\documentclass[prd,preprint,eqsecnum,nofootinbib,amsmath,amssymb,
               tightenlines,dvips]{revtex4}
\usepackage{graphicx}
\usepackage{bm}

\begin {document}


\def\Mrowczynski{Mr\'owczy\'nski}
\def\Qs{Q_{\rm s}}

\def\alphas{\alpha_{\rm s}}

\def\D{{\bm D}}

\def\k{{\bm k}}
\def\half{{\textstyle{\frac12}}}

\def\p{{\bm p}}

\def\x{{\bm x}}

\def\v{{\bm v}}
\def\E{{\bm E}}
\def\B{{\bm B}}
\def\A{{\bm A}}

\def\grad{{\bm\nabla}}

\def\Im{\operatorname{Im}}

\def\lmax{\ell_{\rm max}}
\def\mmax{m_{\rm max}}
\def\ldamp{\ell_{\rm damp}}
\def\mdamp{m_{\rm damp}}

\def\ms{m_{\rm s}}

\def\iso{\theta}   
\def\kmax{k_{\rm max}}
\def\ttau{\tilde\tau}
\def\kunstable{k_{\rm unstable}}
\def\tot{{\rm tot}}

\def\expon{\nu}


\title
    {
    Non-Abelian Plasma Instabilities for Extreme Anisotropy    
    }

\author{Peter Arnold}
\affiliation
    {%
    Department of Physics,
    University of Virginia, P.O. Box 400714
    Charlottesville, Virginia 22904-4714, USA
    }%
\author{Guy D. Moore}
\affiliation
    {%
    Department of Physics,
    McGill University, 3600 rue University,
    Montr\'eal QC H3A 2T8, Canada
    }%

\date {June, 2007}

\begin {abstract}%
    {%
      Thermalization of quark-gluon plasmas in heavy-ion collisions
      is a difficult theoretical problem.
      One theoretical goal has been to understand
      the physics of thermalization in the relatively simplifying limit
      of arbitrarily high energy collisions, where the running
      coupling $\alpha_{\rm s}$ is weak.
      One of the current roadblocks to achieving this goal is
      lack of knowledge about the behavior of plasma instabilities
      when particle distributions are highly anisotropic.
      In particular, it has not been known how the magnetic fields
      generated by plasma instabilities scale with anisotropy.
      In this paper, we use numerical simulations in a first attempt
      to determine this scaling.
    }
\end {abstract}

\maketitle
\thispagestyle {empty}


\section {Introduction and Results}
\label{sec:intro}

How do non-abelian plasmas that start far from equilibrium, such as
quark-gluon plasmas produced in heavy ion collisions,
equilibrate?  This question has proven difficult to answer in detail
even in the theoretical simplifying limit of weak coupling, appropriate
to arbitrarily high energy collisions.  A pathbreaking first attempt was
made by Baier, Mueller, Schiff, and Son \cite{bottom_up}, who analyzed
equilibration in the weak coupling limit via scattering processes of
individual particles.  The resulting picture of quark-gluon plasma
equilibration is known as the bottom-up scenario.  It was later
realized,
however, that collective effects in the form of magnetic
plasma instabilities, known as Weibel or filamentary instabilities,
necessarily play a role in bottom-up equilibration \cite{ALM}.%
\footnote{
  For a sample of earlier discussions of the possible role of
  Weibel instabilities in quark-gluon plasma thermalization,
  see Refs.\ \cite{plasma_old,RS}.
}
These
instabilities have long been known in traditional plasma physics
\cite{weibel}, but their
non-abelian counterparts develop somewhat differently.  The effect of
non-abelian interactions on the late-time development of plasma
instabilities has been studied over the past few years with numerical
simulations \cite{RRS,RRS2,RV,Nara,DNS,BodekerRummukainen,linear1,linear2}.  
A great deal has been learned, but these
simulations have not significantly
explored the extremely non-equilibrium conditions
relevant to the initial phase of bottom-up thermalization.

In particular, Weibel instabilities are generated by anisotropic
distributions of plasma particle momenta, as measured in local plasma
rest frames.  So far, simulations have mostly
focused on the case of moderate anisotropy.%
\footnote{%
    Two exceptions are the paper of B\"odeker and Rummukainen
    \cite{BodekerRummukainen}, with similar methods and aims to the
    current work, and the paper by Dumitru, Nara, and Strickland
    \cite{DNS}, which focuses on an initially perfectly planar
    distribution which is allowed to dynamically broaden with time.
    }
The bottom-up scenario,
however, generates parametrically extreme anisotropies early on,
before thermalization is
achieved.  Thermalization, of course, eventually produces isotropic
(thermal) momentum distributions in local plasma rest frames.  As an
example, in the original bottom-up scenario (ignoring plasma
instabilities), at one particular pre-thermalization moment of the
expansion, the local distribution of particle velocities looks like a
pancake in momentum space, with
\begin {equation}
   p_z \sim g \, p_\perp ,
\label {eq:isotropy0}
\end {equation}
where $g$ is the QCD coupling constant and $z$ is the beam direction.
Formally, in the limit of arbitrarily weak coupling $g$, this represents
extreme anisotropy.  To understand equilibration in the weak coupling
limit, one must therefore understand the development of
plasma instabilities for the case of extreme anisotropy, $p_z/p_\perp \ll 1$.
The purpose of this paper is to make a first attempt to explore this
limit using numerical simulations.

Discussion of weak-coupling thermalization starts from the saturation
picture of high-energy heavy ion collisions, where there is initially
a non-perturbatively large phase-space density $f \sim 1/g^2$ of
low $x$ gluons with momentum of order the saturation scale $\Qs$.
These initial gluons are the ``hard'' particles in discussions of
thermalization.
We formally consider the case where $\Qs$ is so large that
the running coupling $\alphas(\Qs)$ can be treated as arbitrarily small.
Bottom-up thermalization describes what happens as the plasma
subsequently expands
one-dimensionally between the two retreating pancakes of nuclear debris.
The expansion reduces the density of hard particles enough that one
can treat them perturbatively for times $\tau \gg 1/\Qs$.
In the first stage of the original bottom-up picture of thermalization,
which corresponded to $1 \ll \Qs\tau \ll g^{-3}$, the one-dimensional
expansion effectively red-shifts the component $p_z$ of hard particle
momenta along the beam axis, as measured in local plasma rest frames.
For free particles, the expansion would drive the system away from
local anisotropy as
\begin {equation}
  \frac{p_z}{p} \sim (\Qs\tau)^{-1} .
  \qquad\qquad \mbox{(free streaming)}
\label {eq:isotropy1}
\end {equation}
However, small-angle
$2{\to}2$
collisions between the hard particles tend to broaden $p_z/p$,
softening the anisotropy to
\begin {equation}
  \frac{p_z}{p} \sim (\Qs\tau)^{-1/3}
  \qquad\qquad \mbox{(original bottom-up)}
\label {eq:isotropy2}
\end {equation}
in the
original bottom-up analysis of Baier {\it et al.}\ \cite{bottom_up}.
This is a balance between one-dimensional expansion driving the system
away from isotropy and collisions driving it towards isotropy.
In Baier {\it et al.}'s analysis,
this relatively simple state of affairs continues until
parametrically late times $\Qs\tau \sim g^{-3}$, when other interesting
things start to happen to bring about the eventual thermalization of
the plasma.

Plasma instabilities already play a role in the relatively simple
first stage of bottom-up thermalization, however, and we will focus
on this stage to motivate our investigation.
In particular, plasma instabilities provide another mechanism
to drive the system towards isotropy, and they change the exponent
in (\ref{eq:isotropy2}).  Weibel instabilities are associated with
the creation of large, soft magnetic fields, which randomly bend
the directions of the particles.  How much bending occurs depends
on the size of these magnetic fields $B$.
Unfortunately, the parametric size of $B$ in the case of extreme anisotropy
($p_z/p \ll 1$) has not been clear.
As an example, there are two different guesses that have been made in
the literature
\cite{BnewBUP,kminus2}, which would modify the original first-stage bottom-up
behavior (\ref{eq:isotropy2}) to
\begin {equation}
  \frac{p_z}{p} \sim
  \begin {cases}
     (\Qs\tau)^{-1/4}, & \mbox{Ref.\ \cite{BnewBUP};} \\
     (\Qs\tau)^{-1/8}, & \mbox{Ref.\ \cite{kminus2}.}
  \end {cases}
\label {eq:isotropy3}
\end {equation}
The goal of this paper is to make a first attempt at resolving
the issue by measuring the
dependence of the soft magnetic fields $B$, caused by non-abelian Weibel
instabilities, on the anisotropy of the hard particle distribution.


\subsection {Review: The limiting size of unstable magnetic fields}
\label {sec:Bstar}

Let $f_0(\p)$ be the phase-space distribution of particles in the
plasma, so that the density $n$ is
\begin {equation}
   n = \int \frac{d^3p}{(2\pi)^3} \, f_0(\p) .
\label {eq:n}
\end {equation}
For moderately anisotropic $f_0(\p)$, there is a single parametric
scale of soft physics in the plasma which characterizes plasmon masses,
Debye screening, and Weibel instabilities.
For definiteness, we can take the scale of soft physics to be the
effective mass $m_\infty$ of hard gluons in the plasma, given by
\cite{MrowThoma,Boltzmann}
\begin {equation}
   m_\infty^2 \equiv
   g^2 \nu t_R \int \frac{d^3p}{(2\pi)^3} \>
   \frac{f_0(\p)}{p} \,,
\label {eq:minf}
\end {equation}
where there is an implicit sum over species, $t_R$ is a group factor,
and $\nu$ counts the number of non-color
degrees of freedom ({\it e.g.} spin) for a given species.%
\footnote{
   For a plasma of gluons, $\nu = 2$ and $t_R = 3$.
   The $\nu_s$  in Ref.\ \cite{Boltzmann} is this paper's $\nu$ times
   the dimension of the particle's color representation.
}
For moderately anisotropic $f_0$, the typical instability wavenumber
$k_{\rm unstable}$
and growth rate $\gamma$ are both of order $m_\infty$.
Perturbation theory can be used to study the growth of instabilities
from small seed fields.  These instabilities cease to grow when
their magnetic fields become large enough that their non-abelian
self-interaction becomes important and perturbation theory
breaks down \cite{linear1,RRS2}.
Crudely speaking, that happens
when gauge fields become important in soft covariant derivatives
$D = \partial - i g A \sim i (k - gA)$, so that
\begin {align}
  A \sim \frac{k}{g} \sim \frac{m_\infty}{g} ,
  \qquad\qquad &
  \mbox{(moderate anisotropy)}
\\
\intertext{and}
  B_* \sim kA \sim \frac{k^2}{g} \sim \frac{m_\infty^2}{g} .
  \qquad\qquad &
  \mbox{(moderate anisotropy)}
\label {eq:BmaxModerate}
\end {align}
We write $B$ with an asterisk subscript
to denote, roughly speaking,
the limiting size of the magnetic fields associated with
unstable modes.  This excludes other (higher momentum) modes
which are excited at late times, associated with
a cascade of plasmons that we will review later.

\begin{figure}[t]
\includegraphics[scale=0.35]{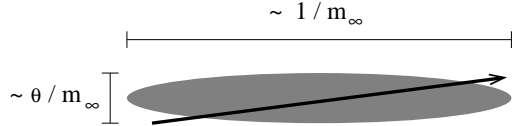}
\caption{%
    \label{fig:coherent}
    A schematic picture of a hard particle crossing a region of
    coherent magnetic field.
}
\end{figure}

For extremely anisotropic distributions $f_0(\p)$, we have an additional
parameter in the problem: the amount of anisotropy.  Motivated by the
application to bottom-up thermalization, we will focus on oblate
distributions that are axi-symmetric about the beam axis $z$,
and we will roughly characterize the amount of anisotropy by
the typical magnitude
\begin {equation}
  \iso \equiv \frac{|p_z|}{p} = |v_z| .
\end {equation}
Since $\iso$ is parametrically small in the first stage of bottom-up
thermalization, we need to know how the physics of instabilities
depends parametrically on $\iso$.
A perturbative analysis of the instability shows that typical unstable modes
have wave numbers $\k$ and growth rates $\gamma$ of order \cite{ALM}
\begin {equation}
  (k_\perp, k_z) \sim (m_\infty, \kmax)
  \sim \Bigl( m_\infty, \frac{m_\infty}{\iso} \Bigr) ,
\end {equation}
\begin {equation}
  \gamma \sim m_\infty ,
\end {equation}
where
\begin {equation}
   \kmax \sim \frac{m_\infty}{\iso}
\label {eq:kmax}
\end {equation}
is the maximum value of $k$ for unstable modes.%
\footnote{
  A way to remember this is as follows.  The physical role of
  the scale $m_\infty$ in the context of instability growth is that
  $1/m_\infty$ is the time scale for currents to build up large enough to have
  important back-reaction on the fields.
  That $\gamma \propto m_\infty$ follows
  immediately.  Currents only build up if particles remain in a coherent
  region of single-sign field for this time scale.
  In time $1/m_\infty$, particles
  travel a transverse distance $1/m_\infty$,
  so $k_\perp \sim m_\infty$; but they only
  travel a $z$ distance of $\sim \theta/m_\infty$,
  so $k_z \sim m_\infty/\theta$.  This
  is illustrated in Fig.\ \ref{fig:coherent}.
}

What has been unclear is the size of the fields
when unstable modes become non-perturbatively large and cease to
grow.
Here is a simple, hand-waving generalization of
(\ref{eq:BmaxModerate}) which reproduces a conjecture made in
Ref.\ \cite{kminus2}.  Soft covariant derivatives
$D_z = \partial_z - i g A_z$ and $D_\perp = \partial_\perp - i g A_\perp$
will become non-perturbative when
\begin {equation}
   A_z \sim \frac{k_z}{g} \sim \frac{\kmax}{g}
   \qquad
   \mbox{and}
   \qquad
   A_\perp \sim \frac{k_\perp}{g} \sim \frac{m_\infty}{g} \,,
\end {equation}
corresponding to magnetic fields
\begin {equation}
   B_\perp \sim (k_\perp A_z ~~ \mbox{or} ~~ k_z A_\perp)
   \sim \frac{\kmax m_\infty}{g}
\end {equation}
and
\begin {equation}
   B_z \sim k_\perp A_\perp \sim \frac{m_\infty^2}{g} \,.
\end {equation}
The transverse fields dominate, with
\begin {equation}
  B_* \sim \frac{\kmax m_\infty}{g} \sim \frac{m_\infty^2}{\iso g} .
\label {eq:Bus}
\end {equation}
Other arguments for the result (\ref{eq:Bus})
can be found in Ref.\ \cite{kminus2}.%
\footnote{
  See specifically Sec.\ V of Ref.\ \cite{kminus2}.
  Readers of other sections of Ref.\ \cite{kminus2} should be
  aware that some of the arguments there may be overly simplistic.
  In particular, see Ref.\ \cite{MSW} for related discussion.
}
In contrast, an earlier discussion by Ref.\ \cite{BnewBUP} assumed
that $B_* \sim m_\infty^2/g$ as in (\ref{eq:BmaxModerate}).
One of our goals will be to distinguish these two possibilities using
simulations to investigate the exponent $\expon$ in
\begin {equation}
  B_* \sim \frac{m_\infty^2}{\iso^\expon g} \,,
\label {eq:Bn}
\end {equation}
where
\begin {equation}
  \expon =
  \begin {cases}
     0, & \mbox{Ref.\ \cite{BnewBUP};} \\
     1, & \mbox{Ref.\ \cite{kminus2};} \\
     2, & \mbox{Nielsen-Olesen limit.}
  \end {cases}
\label {eq:expon}
\end {equation}
Here we show a third magnetic scale for comparison, 
the Nielsen-Olesen limit.
It is associated with Nielsen-Olesen instabilities
and is discussed in Appendix \ref{app:NO}.

As in Refs.\ \cite{BnewBUP,kminus2}, one can use (\ref{eq:Bn}) to
parametrically
determine how particle scattering from these fields
broadens $p_z/p$, and balance this against the one-dimensional
expansion to determine how $p_z/p$ scales with time.  In a chaotic
system, the coherence length of the unstable magnetic fields will be
of order their wavelength, and so be of order
\begin {equation}
  l_\perp \sim \frac{1}{k_\perp} \sim \frac{1}{m_\infty}
  \qquad \mbox{and}\qquad
  l_z \sim \frac{1}{k_z} \sim \frac{\iso}{m_\infty}
\end {equation}
in the transverse and $z$ directions, as depicted qualitatively
in Fig.\ \ref{fig:coherent}.
The particles have velocity $(v_\perp,v_z) \sim (1,\iso)$ and will take
time $\delta t \sim 1/m_\infty$ to cross such a region.
In that time, the magnetic force ${\bm F}$ will change the particle's
$p_z$ by
$\delta p_z \sim F_z t \sim g B_* l_\perp \sim g B_*/m_\infty$.
In time $\tau$, the particle will random walk through $N \sim \tau/l_\perp$
such changes, giving a total change of order
\begin {equation}
  \Delta p_z \sim N^{1/2} \, \delta p_z
  \sim (m_\infty \tau)^{1/2} \frac{gB_*}{m_\infty}
  \sim \frac{(m_\infty^3\tau)^{1/2}}{\iso^\expon} \,.
\label {eq:DeltaP}
\end {equation}
This will broaden the particle
distribution to
\begin {equation}
  \iso \equiv \frac{p_z}{p} \sim \frac{\Delta p_z}{p}
  \sim \frac{(m_\infty^3\tau)^{1/2}}{\iso^\expon \Qs} \,.
\end {equation}
Solving self-consistently for $\iso$,
\begin {equation}
  \iso \sim
  \left(\frac{(m_\infty^3\tau)^{1/2}}{\Qs}\right)^{1/(1+\expon)}.
\label {eq:iso}
\end {equation}
Now we just need to know how $m_\infty$ depends on time.
This was determined for the first stage of bottom-up thermalization
by very simple arguments in the original work of Baier {\it et al.}
\cite{bottom_up} and remains unchanged in the presence of plasma
instabilities.
Comparing (\ref{eq:n}) and (\ref{eq:minf}), one sees that
$m_\infty^2 \sim g^2 n/p \sim g^2 n/\Qs$.
Initially, at saturation, $n \sim \Qs^3/g^2$.
In the first stage of bottom-up, there is no significant change in the
number of hard particles, and so hard particle number density $n$
dilutes from this initial value by the scale
factor $\Qs\tau$ of one-dimensional expansion, so that
$n \sim \Qs^3/g^2(\Qs\tau)$.  Putting everything together,
\begin {equation}
  m_\infty \sim \tau^{-1/2} \Qs^{1/2} .
\label {eq:minfBup}
\end {equation}
Inserting this into (\ref{eq:iso}) produces the scaling (\ref{eq:isotropy3})
of $p_z/p$ with time quoted in the introduction.

Table \ref{tab:bottom_up} summarizes a variety of weak-coupling predictions for
the first phase of the original bottom-up scenario \cite{bottom_up} as
well as its modification due to instabilities as conjectured
in Ref.\ \cite{kminus2}, corresponding to the limiting field
(\ref{eq:Bus}) above.
Here, we have defined the dimensionless time $\ttau \equiv \Qs\tau$.
Since many readers may be more familiar with equilibrium plasma physics
than with the scales of bottom-up thermalization, we also show, for the
sake of qualitative comparison, what similar scales would be for
(i) an equilibrium plasma at temperature $T$, and (ii) a ``squashed''
equilibrium plasma which has the same density $n \sim T^3$ and typical
energy $p \sim T$ of particles
but has particle momenta distributed anisotropically with
$p_z/p \ll 1$.  In the thermal case, the hierarchy of different mass scales
is controlled by the small parameter $g$.  In the bottom-up case, it
is instead controlled by the small parameter $(\Qs\tau)^{-1}$.
Note that the bottom-up scales satisfy the hierarchy that the
instability growth rate is parametrically faster than both the
expansion rate%
\footnote{
   See Ref.\ \cite{HEL} for a recent analysis of the unfavorable effects
   of expansion on instabilities for heavy ion collisions at
   realistic (rather than arbitrarily large) energies.
}
and the rate for individual (incoherent) $2{\to}2$
hard particle collisions.  For the purpose of determining the
limiting $B_*$ of (\ref{eq:Bn}) in the specific
context of bottom-up thermalization
in the weak coupling limit,
this hierarchy allows one to ignore the effects of
both expansion and individual collisions when simulating plasma
instabilities \cite {ALM}.

\begin {table}
\begin{tabular}{|l|c|cc|cc|}
\hline
& general
& thermal & squashed & original  & guess \\
&
&         & thermal  & bottom-up & Ref.\ \cite{kminus2} \\[5pt]
\hline
hard particle momenta $p$ &
  &
  $T$ & $T$ & $\Qs$ & no change \\
particle isotropy $\iso{\equiv}p_z/p$ &
  &
  1 & $\iso \ll 1$ & $\ttau^{-1/3}$ & $\ttau^{-1/8}$ \\
hard particle density $n$ &
  &
  $T^3$ & $T^3$ & $\ttau^{-1} \Qs^3/g^2$ & no change \\
phase space density $f$ &
  $n/\iso p^3$ &
  1  &  $\iso^{-1}$  & $\ttau^{-2/3}/g^2$ & $\ttau^{-7/8}/g^2$ \\
hard plasmon mass $m_\infty$ &
  $\sqrt{g^2 n/p}$ &
  $gT$ & $gT$ & $\ttau^{-1/2} \Qs$ & no change \\
particle collision rate &
  $g^4 n (1{+}f) / m_\infty^2$ &
  $g^2 T$ & $g^2 T/\theta$ & $\ttau^{-2/3} \Qs$ & $\ttau^{-7/8} \Qs$ \\
expansion rate &
  &
  -- & -- & $\tau^{-1}$ & no change \\
\hline
instability wave number &
  $m_\infty/\iso~{}^{(*)}$ &
  -- & $gT/\iso$ & $\ttau^{-1/6} \Qs$ & $\ttau^{-3/8} \Qs$ \\
instability growth rate &
  $m_\infty~{}^{(*)}$ &
  -- & $gT$ & $\ttau^{-1/2} \Qs$ & no change \\
\hline
\end{tabular}
\label {tab:bottom_up}
\caption{%
  A table of the parametric dependence of various scales for
  (i) a thermal distribution,
  (ii) a ``squashed'' thermal distribution with the same density $n$ but
  extreme momentum anisotropy $\iso$,
  (iii) the first stage ($1 \ll \Qs\tau \ll g^{-3}$) of the
  original bottom-up thermalization scenario of
  Baier {\it et al.}\ \cite{bottom_up},
  and (iv) the changes to bottom-up due to instabilities based on the
  conjectured dynamics of Ref.\ \cite{kminus2}.
  In this table, $\ttau \equiv \Qs\tau$ and the phase space density $f$
  refers to the largest values of $f(\p)$ (and not to the angular-averaged
  values).
  The ``particle collision rate'' refers to the rate of individual
  (incoherent), small-angle $2 \to 2$ scattering of hard particles
  from each other.  The instability wave number refers to $k \sim k_z$.
  In contrast, $k_\perp \sim m_\infty$ as discussed in the text.
  An asterisk $^{(*)}$ indicates general formulas that apply only to
  moderate to extreme anisotropy and not to isotropic or nearly
  isotropic situations.
}%
\end {table}


\subsection {Overview of simulation method and what we measure}
\label {sec:measure}

In order to cleanly separate scales in the weak-coupling limit,
simulations are carried out for the hard-loop effective theory
of soft excitations as in Refs.\ \cite{MRS,linear1,RRS,RRS2}.
This effective theory is a non-abelian version of the linearized
Vlasov equations of traditional plasma physics, which are based on
collisionless kinetic theory for hard particles coupled to a soft,
classical gauge field.
We use the formulation of Ref.\ \cite{linear2}, where the equations
are
\begin {subequations}
\label{eq:vlasov}
\begin{eqnarray}
\label{eq:YangMills}
D_\nu F^{\mu \nu}(\x,t) & = & \int_\v v^\mu W(\v,\x,t) \, , \\
(D_t + \v \cdot \D_\x) W & = & m_\infty^2 \left[ \E \cdot (2\v-\grad_\v)
	+ \B \cdot ( \v \times \grad_\v ) \right] \Omega(\v) \, .
\label{eq:W}
\end{eqnarray}
\end {subequations}
Here, the field $W^a(\v,\x)$ represents the
net (adjoint) color of all particles moving in direction $\v$ at point
$\x$.
The first equation is the Yang-Mills field equation, with $W(\v)$
giving rise to a current.
The second equation, derived in this form in
Ref.\ \cite{linear1}, shows how electric and
magnetic fields can polarize the colorless distribution of particles to
create a net color moving in each direction.
In this equation, $\Omega(\v)$ is a static quantity which parametrizes
the angular distribution of the initial, background distribution
$f_0(\p)$ of hard particles.  $W(\v,\x,t)$ is generated by
small fluctuations of $f(\p)$ from $f_0(\p)$.
(For the particular weak-coupling questions treated here,
it is allowable to treat the hard particle fluctuations as small.)
The dynamics of the soft fields is equivalent to
that of hard-loop effective theory \cite{MRS}.
The use of classical equations (\ref{eq:vlasov}) can be justified for
the applications at hand because deBroglie wavelengths of the hard
particles are parametrically small compared to the soft physics
distance scales, and because the instability causes the soft gauge
fields to grow parametrically large enough to be classical.

To discretize these equations for simulation, we follow the methods
of Refs.\ \cite{linear1,linear2}, with a small but important change
discussed in Sec.\ \ref{sec:simulate} to allow us to more efficiently
simulate the case of extremely anisotropic distributions.
Also, like previous studies of Weibel instabilities, all of our simulations
will be for SU(2) gauge theory for reasons of computational simplicity.
We expect this to be qualitatively
similar to SU(3) gauge theory; we are not aware of any
reason why they would be different.

\begin{figure}[t]
\includegraphics[scale=0.5]{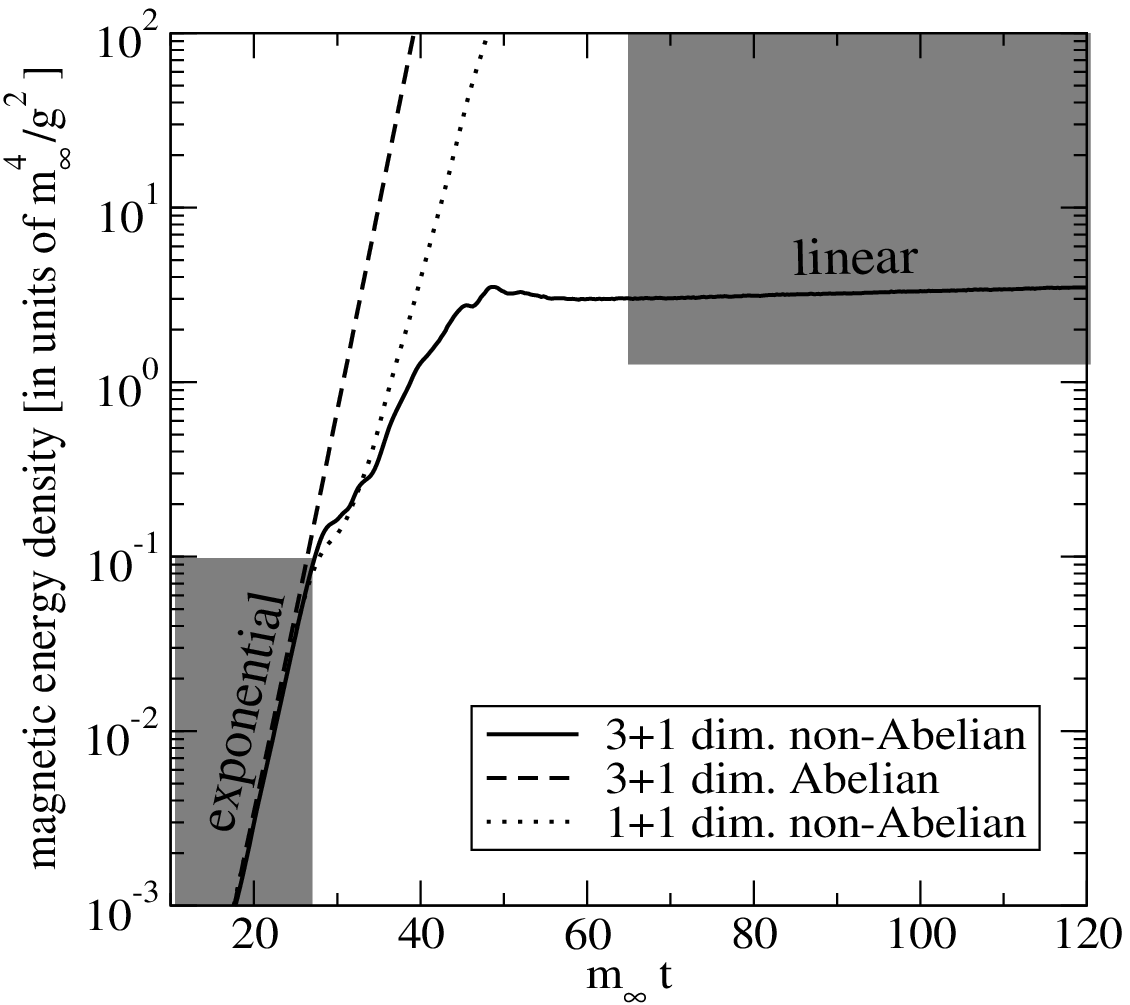}
\includegraphics[scale=0.5]{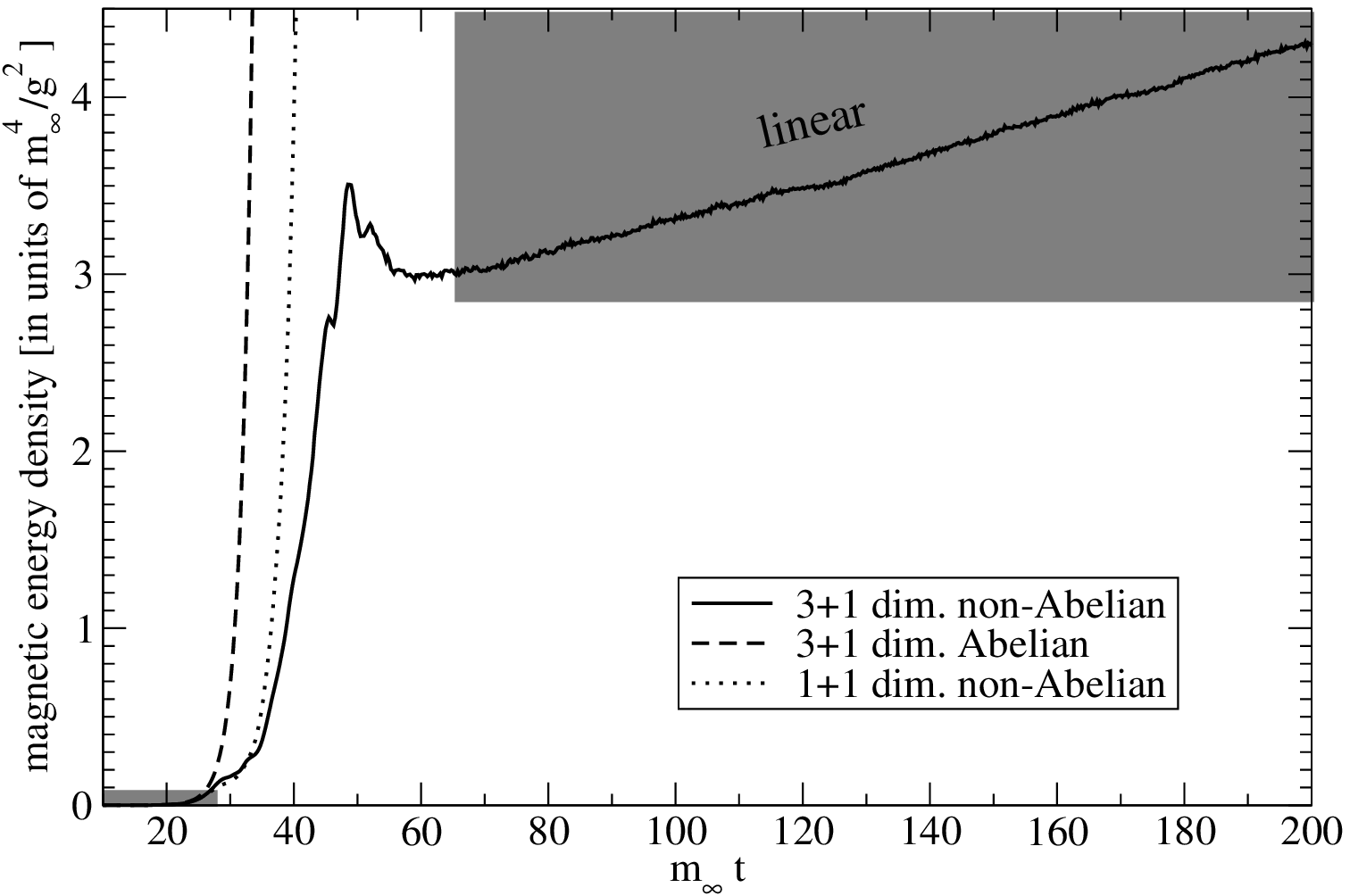}
\caption{%
    \label{fig:linear1}
    Magnetic energy vs.\ time for a sample simulation from Ref.\
    \cite{linear1} for moderate anisotropy, starting from a very
    small seed for the Weibel instability.  The figures are the same
    except that the vertical axis is
    logarithmic in the left-hand figure and linear in the
    right-hand figure.
    The solid line is the result for non-abelian gauge theory in
    three spatial dimensions.
    For comparison, the dashed line shows a simulation in an
    abelian theory, and the dotted line shows a non-abelian simulation
    in one spatial dimension.
}
\end{figure}

Fig.\ \ref{fig:linear1} shows an example from Ref.\ \cite{linear1},
showing the total energy density in soft magnetic fields as a function of
time.
This particular simulation was for moderate anisotropy and started from
tiny initial conditions for the gauge fields.
There is exponential growth at early times, due to the
instability, and linear growth
at late times.  The linear growth does not represent continued growth
of the unstable modes.  Instead, the unstable modes stop growing but,
through interactions, pump energy into a cascade of increasingly
higher momentum, stable modes \cite{linear2}.
This cascade takes
the form of a gas of plasma excitations of the classical gauge field with
momentum $q \gtrsim \kunstable$,
which are perturbative for $q \gg \kunstable$.
At late times, the total classical magnetic
field energy density
${\cal E}^B_\tot = \half B^2$
is dominated by the energy of these perturbative
plasma excitations, rather than the energy density
${\cal E}^B_* \sim \half B_*^2$
of the softer
($k\sim\kunstable$) unstable modes.  For this reason, we cannot simply
measure the total magnetic energy density $\half B^2$
at late times and take
a square root to find the limiting size $B_*$ of the magnetic fields associated
with unstable modes.
And it is the soft fields $k \sim \kunstable$,
not the higher momentum plasmon excitations, which
dominate the scattering of hard particles and so determine the
evolution (\ref{eq:Bn}) relevant to bottom-up thermalization
\cite{kminus2}:
Even though the $k \sim \kunstable$ modes carry less energy, they are
more effective at scattering.
To determine what we really want to know, we need some
measurement other than the total magnetic energy density
at a single late time.

In this paper,
we will use an indirect method to investigate anisotropy dependence
which is relatively easy to implement.
We will  measure the slope $d{\cal E}^B_\tot/dt$
of the late-time
linear growth of the total magnetic energy density
and determine how it scales with
anisotropy.  Here is a model of how one might expect this slope to
behave.
The source of increasing total magnetic energy comes from the unstable
modes, which take energy from the hard particles and, through
interactions, dump it into the cascade of plasmons.
As a thought experiment, imagine that half the energy density ${\cal E}^B_*$
in the unstable
modes were abruptly transferred to the cascade of plasmons.  How long
would it take the unstable modes to grow back to their original,
limiting size?  Parametrically, the time should be of order
the inverse instability growth rate $t \sim \gamma^{-1} \sim m_\infty^{-1}$.
(Even though this is a perturbative estimate of the growth rate,
it should still be parametrically correct in the region where
perturbation theory starts to break down.)  So the rate energy is
pumped into the cascade can be expected to be of order $\gamma {\cal E}^B_*$:
\begin {equation}
   \frac{d {\cal E}^B_\tot }{dt} \equiv
   \frac{d}{dt} \left( \half B^2 \right)_\tot
   \sim \gamma {\cal E}^B_*
   \sim \gamma B_*^2
   \sim \frac{m_\infty^5}{\iso^{2\expon} g^2} ,
\label {eq:dEdtn}
\end {equation}
where we have used the parametrization (\ref{eq:Bn}) of $B_*$.
By measuring how this slope scales with $\iso$, we can extract
the desired exponent $\expon$ that determines $B_*$, assuming that the
physical argument for (\ref{eq:dEdtn}) is correct.

Following Ref.\ \cite{linear2}, we will generally start our simulations
with large initial gauge fields so that we can quickly and
easily get to the late-time limiting behavior.
For extremely anisotropic hard particle distributions,
this is a non-trivial choice: recent simulations
\cite{BodekerRummukainen}
starting instead from
tiny initial gauge fields find qualitatively different
behavior.  We will return to this point in Sec.\ \ref{sec:small},
where we argue that large initial conditions are appropriate
to understanding how bottom-up thermalization
is modified by instabilities.

We should note that the limiting field $B_*$ we have used to present
a qualitative picture of the physics of instabilities is not a
precisely defined quantity.
Unlike the total magnetic field $B_{\rm tot}$, we know of no
unique, convention-independent, gauge-invariant definition of the
magnitude of $B_*$, and so $B_*$ is only useful for parametric
estimates.
In contrast, the observable
(\ref{eq:dEdtn}) discussed above is gauge-invariant.


\subsection {Overview of Results}

\begin{figure}[t]
\includegraphics[scale=0.35]{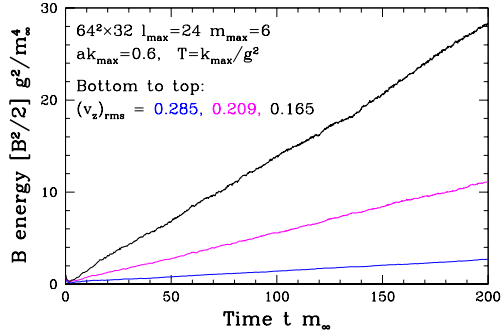}
\caption{%
    \label{fig:dEdt_example}
    Examples of the linear growth of total magnetic energy
    with time.  From bottom to top, the three curves have 
    $\kmax/m_\infty = 2.16$, $3.15$, and $4.10$ respectively
    (corresponding to the $N_\Omega=3$, 5, and 7 distributions
    described in Sec.\ \ref{sec:fguy}).
}
\end{figure}

Fig.\ \ref{fig:dEdt_example} shows an example of
the total magnetic energy density
${\cal E}^B_\tot$ vs.\ time for three different anisotropies, starting
from strong, non-perturbative initial conditions which we shall
detail later.  One way to parametrize the amount of anisotropy
is to rewrite (\ref{eq:kmax}) as
\begin {equation}
   \theta \sim \frac{m_\infty}{\kmax} \,,
\end {equation}
where we compute $m_\infty/\kmax$ perturbatively for each
background hard particle velocity distribution $\Omega(\v)$
[i.e.\ each distribution $f_0(\p)$] that we simulate.
Fig.\ \ref{fig:dEdt_example} shows increasing slope for
increasing anisotropy.
We will then rewrite the scaling form (\ref{eq:dEdtn}) in
the form
\begin {equation}
   \frac{g^2 \, d {\cal E}^B_\tot / dt}
        {m_\infty^{4-2\expon} \kmax^{2\expon} \gamma_*}
   \sim \mbox{constant} ,
\label {eq:dEdt_kmax}
\end {equation}
where we take $\gamma_*$ to be the largest unstable mode growth
rate computed in perturbation theory.  (This rate approaches
$m_\infty/\sqrt{2}$
in the limit of extreme anisotropy \cite{ALM}, but we
have chosen to keep $\gamma_*$ explicit in our formula
because the approach to this limit is a bit slow.
Details will be given in Sec.\ \ref{sec:fguy}.)

\begin{figure}[t]
\includegraphics[scale=0.35]{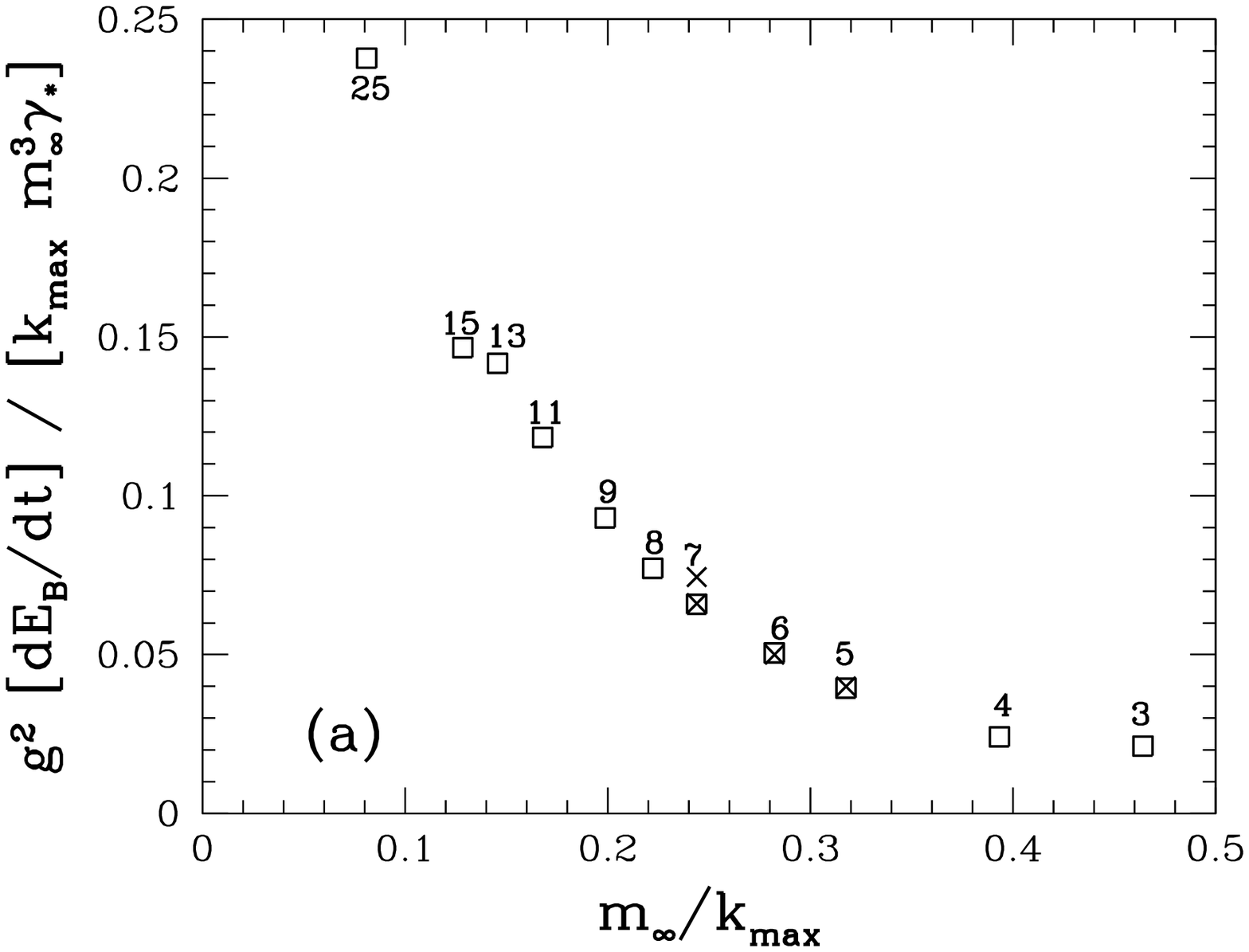} \kern 10pt
\includegraphics[scale=0.35]{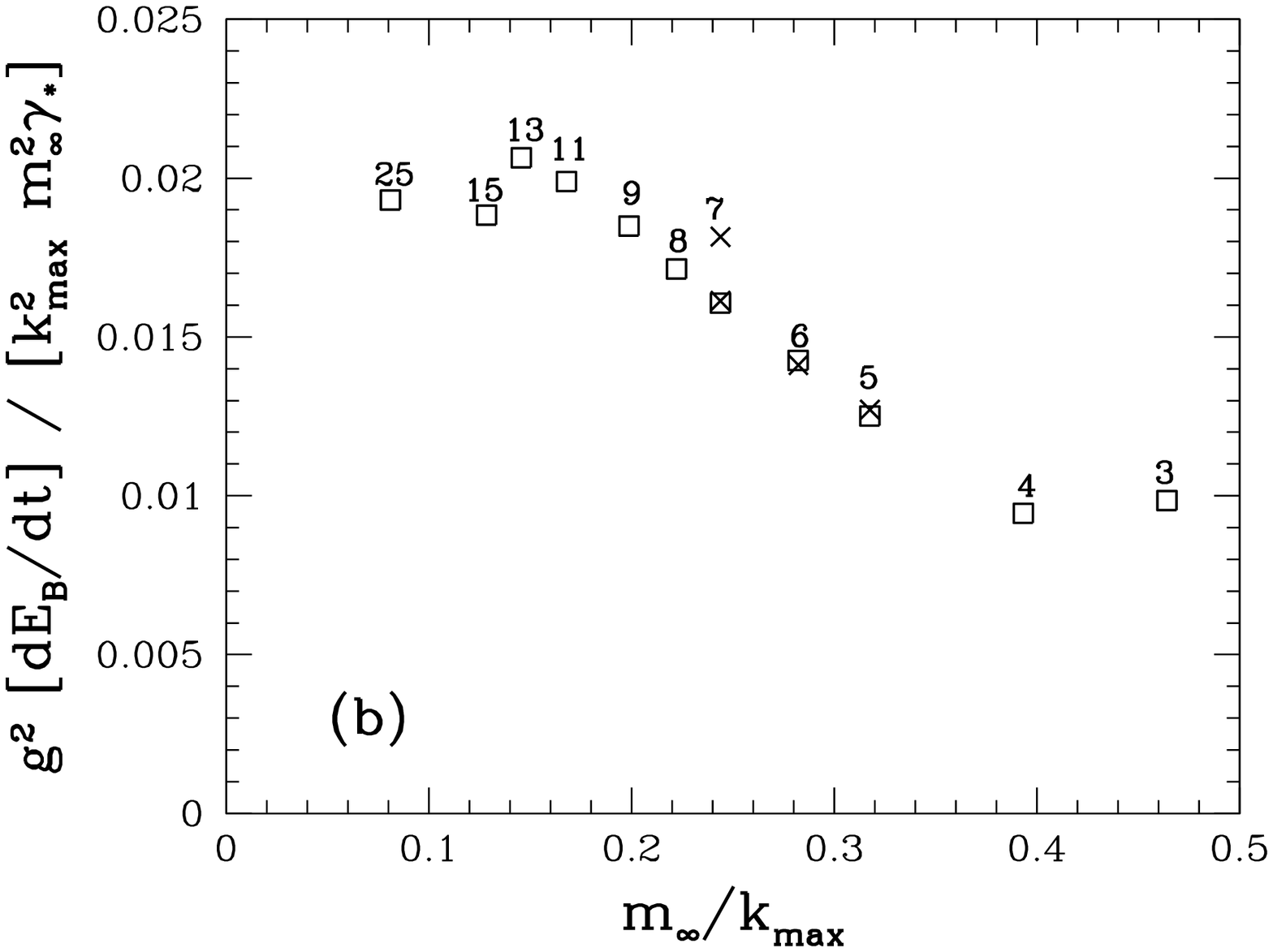} \\[10pt]
\includegraphics[scale=0.35]{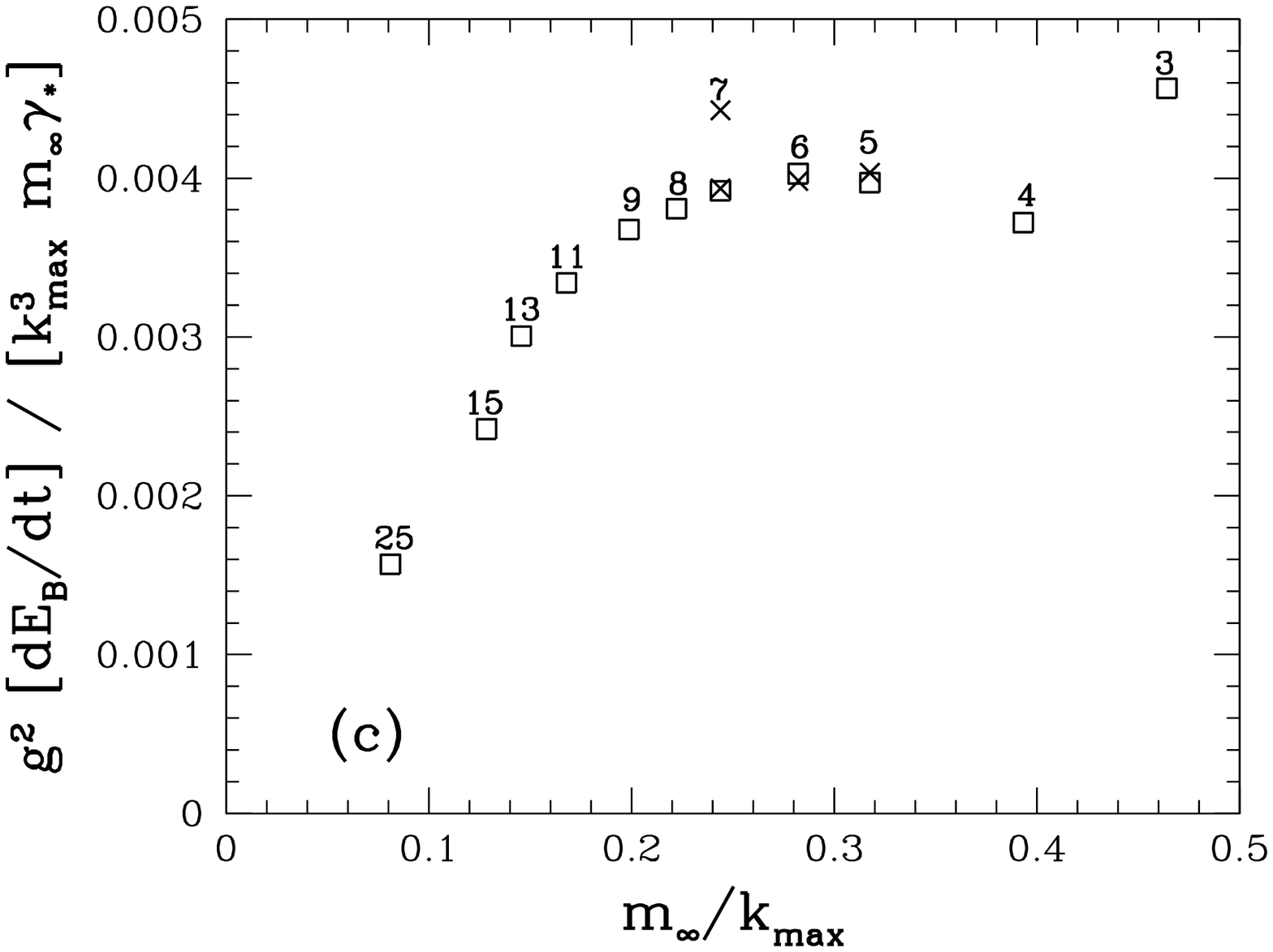}
\caption{%
    \label{fig:dEdt}
    The slope $d{\cal E}^B_\tot/dt$ of the linear growth of total
    magnetic field energy, measured in units of
    $m_\infty^{4-2\nu} \kmax^{2\nu} \gamma_*/g^2$, as a function
    of anisotropy for (a) $\nu=\half$, (b) $\nu=1$, and (c) $\nu=\tfrac32$.
    Simulation parameters are listed in Table \ref{tab:params}
    in Section \ref{sec:simulate}; squares are the default values
    and crosses are alternate values at the bottom of the table.
    The numbers by the data points indicate the order $N_\Omega$ of the
    distribution, as described in Sec.\ \ref{sec:fguy}.
}
\end{figure}

Fig.\ \ref{fig:dEdt} shows the left-hand side of (\ref{eq:dEdt_kmax})
vs.\ our measure $m_\infty/\kmax$ of anisotropy for a variety of
different simulations, plotted with exponents $\nu = \half$, $1$, and
$\tfrac32$.  Each point has systematic errors of order 15\%.
The $\nu=1$ version plausibly approaches a constant in
the extreme anisotropy limit $m_\infty/\kmax \to 0$.
The $\nu=\half$ and $\nu=\tfrac32$ figures clearly rule out
$\nu \le \half$ and $\nu \ge \tfrac32$.
Of the three possibilities $\nu=0$, $1$, and $2$ considered
in (\ref{eq:expon}), we conclude that only $\nu=1$ is consistent
with this measurement.

\section {Large vs.\ small initial conditions}
\label {sec:small}

\begin{figure}[t]
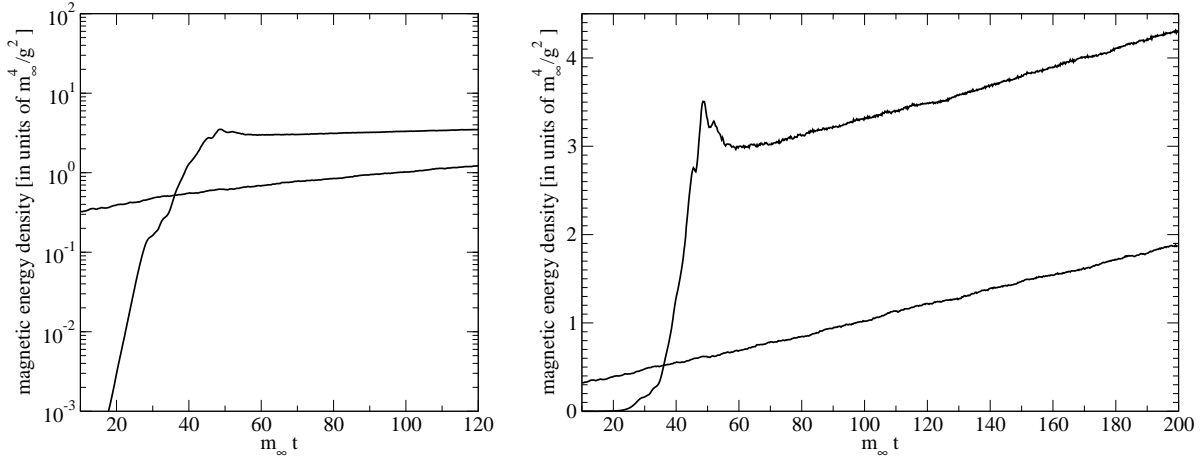

\includegraphics[scale=0.35]{large_v_small.eps} \kern 10pt
\includegraphics[scale=0.35]{large_v_small_linear.eps}
\caption{%
    \label{fig:large_v_small}
    The difference between tiny initial conditions, as in
    Fig.\ \ref{fig:linear1}, and large initial conditions, similar
    to Ref.\ \cite{linear2}.
    (The large initial conditions were set as in Sec.\ \ref{sec:strong}
    but with $T = 4 m_\infty / 3 g^2$, $k_{\rm smear}=2m_\infty$,
    and squeeze $s=1.5$.)
}
\end{figure}

We have initialized our simulations with large initial
gauge fields.
In contrast, many simulations in the past,
such as Fig.\ \ref{fig:linear1},
have started from tiny initial gauge
fields in order to observe the crossover from perturbative,
exponential growth of instabilities to the limiting late-time
behavior.
Fig.\ \ref{fig:large_v_small} shows the difference
for moderate anisotropy: we have superposed
the tiny initial condition simulation of Fig.\ \ref{fig:linear1}
(Ref.\ \cite{linear1}) with an otherwise identical simulation
starting from large intitial conditions.
For tiny initial conditions,
there is a significant spurt of continued exponential-like
growth even after the field strength reaches non-perturbatively
large values.  It is only later, at much higher energy, that
linear growth finally sets in.
B\"odeker and Rummukainen \cite{BodekerRummukainen} have found that
this spurt of post-non-perturbative exponential growth for
tiny initial conditions becomes much more significant for
extreme anisotropy.  In their simulations for extreme
anisotropy, they see only exponential-like growth at late times;
they do not see late-time linear behavior at all.
It is possible that the late-time behavior is ultimately
linear but sets in
at such large field energy that their simulations cannot
reproduce it because of
lattice spacing artifacts.  But regardless, the full story of the
development of plasma instabilities appears to be qualitatively
different depending on whether or not one starts with large or
tiny initial conditions.

\begin{figure}[t]
\includegraphics[scale=0.50]{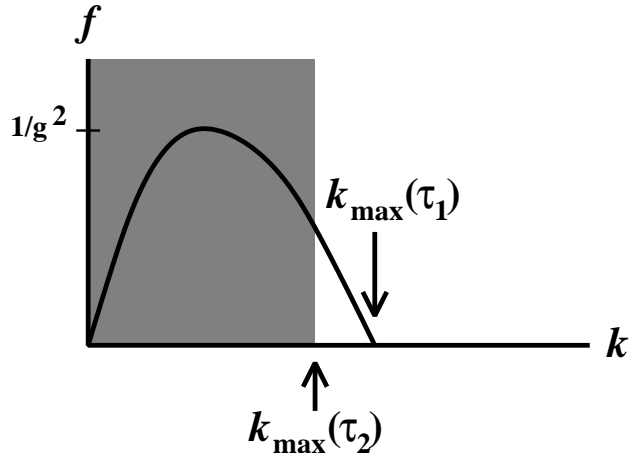}
\caption{%
    \label{fig:seeding1}
    A cartoon of the occupation number $f(k)$ of unstable modes at
    time $\tau_1$, with the shaded area depicting those modes which
    are perturbatively unstable at the later time $\tau_2 = 2 \tau_1$.
}
\end{figure}

Which type of initial condition is relevant to a new
scenario of bottom-up thermalization?
We argue that it is the case of non-perturbatively large
initial conditions that we have investigated in this paper.
Consider some time $\tau_1$ in the first stage of a new bottom-up
scenario that accounts for instabilities, with $\Qs \tau_1 \gg 1$.
We've already reviewed how the instability growth rate is large
compared to the expansion rate $1/\tau_1$, and so
the unstable modes will have grown to become non-perturbatively
big (or perhaps larger).
This population of unstable modes
is depicted very crudely by the curve
in the cartoon of Fig.\ \ref{fig:seeding1}.
(In addition, higher momentum modes may be populated due to
having been unstable at earlier times, or due to interactions.)
It's important to note that the unstable modes will grow to
non-perturbative size
regardless of how small the
initial seed fields for those unstable modes are, because
the instability growth rate is fast and quantum fluctuations
will seed the unstable modes even if nothing else does.%
\footnote{
  In more detail, quantum fluctuations by themselves would correspond
  to fluctuations of order $A \sim \kmax$ in the size of typical
  unstable modes.
  The instability would cause these to grow to non-perturbative
  size $A \sim \kmax/g$ in time of order $\gamma^{-1}$ times the
  log of the size ratio: $\gamma^{-1} \ln(1/g) \sim m_\infty^{-1} \ln(1/g)
  \sim (\tau/\Qs)^{1/2} \ln(1/g)$.
  This time is much shorter than the life $\tau$ of the system
  when $\Qs \tau \gg \ln^2(1/g)$, which we can roughly think of
  as the condition $\Qs \tau \gg 1$ for the applicability of
  bottom-up thermalization, since we have generally not tried to
  keep track of logarithms in discussions of scales.
}

Now consider what happens a little later, at time
$\tau_2 \equiv 2 \tau_1$.  The set of unstable modes shrinks
a bit in $k$ space.  Specifically, combining
(\ref{eq:kmax}), (\ref{eq:iso}), and (\ref{eq:minfBup}), we have
\begin {equation}
   \kmax \sim
   \frac{m_\infty}{\iso}
   \sim \Qs \, (\Qs\tau)^{-(1+2\expon)/[4(1+\expon)]} .
\end {equation}
For definiteness, consider $\expon=1$, for which
$\kmax \sim \Qs \, (\Qs\tau)^{-3/8}$.
Then $\kmax$ decreases
by a factor of $2^{-3/8} \simeq 0.77$
when time $\tau$ increases by a factor of 2,
The unstable modes at the later time $\tau_2$ are therefore shown by
the shaded area in Fig.\ \ref{fig:seeding1}, and we see that they are
already initialized with non-perturbatively large fields because of
the earlier instability growth at time $\tau_1$.

We have tried to make our argument general: Regardless of whether
we start with large or small seeds for instability growth at
$\tau=\tau_1$, we will then get large seeds for instability growth at
later times such as $\tau_2 = 2 \tau_1$.
But the same argument means that we had large seeds at $\tau_1$ also
because of yet earlier instability growth at time $\tau_1/2$.  We can follow
this argument all the way back to times of order the saturation time,
when all the relevant modes of the fields started non-perturbatively
large.
We conclude that the typical unstable modes have high
occupancies at all times during the initial stage of the new bottom-up
scenario.


\section {Simulating extreme anisotropy}
\label {sec:simulate}

To simulate the non-abelian Vlasov equations (\ref{eq:vlasov}), we
need to discretize the arguments $\x$ and $\v$ of the fields
$W(\x,\v,t)$ and $A(\x,t)$.
For $\x$, we put the system on a spatial lattice.
For velocity $\v$, we follow
Refs.\ \cite{BMR,linear1,linear2} and expand in spherical
harmonics $Y_{\ell m}(\v)$, truncating the expansion at some maximum value
$\lmax$ of $\ell$:
\begin {equation}
   W(\x,\v,t) =
   \sum_{\ell=0}^{\lmax} \sum_{m} W_{\ell m}(\x,t) \, \hat Y_{\ell m}(\v) ,
\label {eq:Wx0}
\end {equation}
where our convention is to normalize the spherical harmonics so that
the angular {\it average}\/ of $\hat Y_{\ell m}(\v) \, \hat Y_{\ell' m'}(\v)$
is $\delta_{\ell\ell'} \delta_{m m'}$. 
(We place the hat over $\hat Y_{\ell m}$ as a reminder of this
non-standard normalization convention.)
The axi-symmetric, hard particle background velocity distribution
$\Omega(\v)$ has the expansion
\begin {equation}
   \Omega(\v)
   =\sum_{\ell=0}^{\lmax} \Omega_\ell \, \hat Y_{\ell 0}(\v)
   =\sum_{\ell=0}^{\lmax} \Omega_\ell (2\ell+1)^{1/2} \, P_\ell(v_z) ,
\end {equation}
where the $P_\ell(x)$ are Legendre polynomials.
The explicit form of the equations of motion (\ref{eq:vlasov}) in terms of
the $W_{\ell m}$'s is given in Ref.\ \cite{linear1}.

In practice, we must choose $\lmax$ large enough to obtain results close
to the $\lmax\to\infty$ limit.  More anisotropic distributions
$\Omega(\v)$ will require larger $\lmax$ and therefore greater
computational resources (both memory and time, to store and evolve
more $W_{\ell m}$'s).
Below, we first describe our choice of distributions $\Omega(\v)$
to simulate.  Then we explain and justify our method for
making simulations of very anisotropic distributions practical,
which is to reduce the number of $W_{\ell m}$'s by limiting
$m$ to $|m| \le \mmax$ with $\mmax \simeq 6$.


\subsection {Choice of hard particle distribution \boldmath$f_0(\p)$}
\label{sec:fguy}

For a given maximum $\ell$, we would like to find a velocity
distribution $\Omega(\v)$ of hard particles which is as anisotropic
$(\iso \sim |v_z| \ll 1)$ as possible.
To be physical, $\Omega(\v)$ should be non-negative.%
\footnote{
  We do not know if there would be any problem for simulations if
  $\Omega(v_z)$ had tiny negative values for some $v_z$, but
  it seems safer to avoid this.
}
After some experimentation, we settled on the following form,
parametrized by an integer order $N_\Omega$:
\begin {equation}
   \Omega(v_z) =
      \begin{cases}
         {\cal N}
         (1-v_z^2)\phantom{{}^2} \prod_{i=1}^n (\alpha_i - v_z^2)^2 ,
           & \mbox{for $N_\Omega = 2n+1$;} \\
         {\cal N}
         (1-v_z^2)^2 \prod_{i=1}^n (\alpha_i - v_z^2)^2 ,
           & \mbox{for $N_\Omega = 2n+2$;}
       \end {cases}
\end {equation}
where the normalization ${\cal N}$ is chosen to satisfy our
convention that the angular
average $\Omega_0$ of $\Omega(\v)$ is one.
We choose the $\alpha_i$ to minimize $\langle v_z^2 \rangle$,
performing the minimization numerically for each $N_\Omega$.

The expansion of $\Omega(\v)$ involves spherical harmonics with
$l \le L_\Omega \equiv 2 N_\Omega$, and the corresponding
coefficients $\Omega_l$ are listed in Appendix \ref{app:Omega}
for various choices of $N_\Omega$.  We will see later, looking at
the $\lmax$ dependence of results, that we can get close to the 
$\lmax \to\infty$ limit for a given $N_\Omega$ using
$\lmax \gtrsim 1.5\, L_\Omega = 3 N_\Omega$.

\begin{figure}[t]
\includegraphics[scale=1.00]{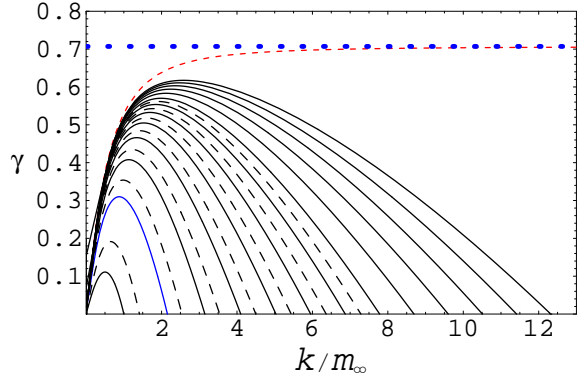}
\caption{%
    \label {fig:gamma_vs_k}
    Perturbative instability growth rates $\gamma(k)$ vs.\ $k/m_\infty$
    for the values of $N_\Omega$ listed in Table \ref{tab:fguy}.
    Solid (dashed) black lines are odd (even) $N_\Omega$, staring with
    $N_\Omega=1$ at the bottom and running up to $N_\Omega=25$ at the top.
    The horizontal dotted line is the maximum possible
    $\gamma$, which is $1/\sqrt2$, and the dashed line
    approaching it is the case $N_\Omega=\infty$, given in Ref.\ \cite{ALM}.
    }
\end{figure}

\begin{figure}[t]
\includegraphics[scale=0.30]{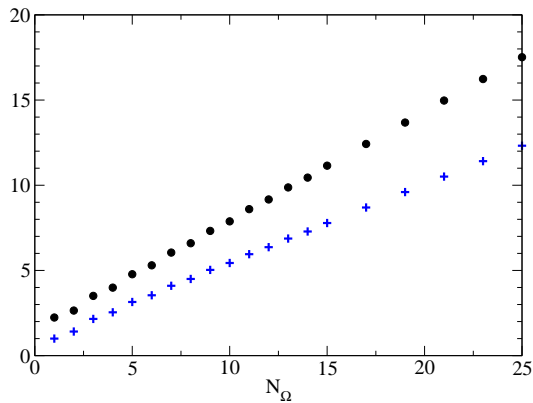}
\caption{%
    \label{fig:vz_vs_n}
    The values of $1/(v_z)_{\rm rms}$ (circles) and
    $\kmax/m_\infty$ (crosses) plotted vs.\ $N_\Omega$.
    }
\end{figure}

\begin{figure}[t]
\includegraphics[scale=0.30]{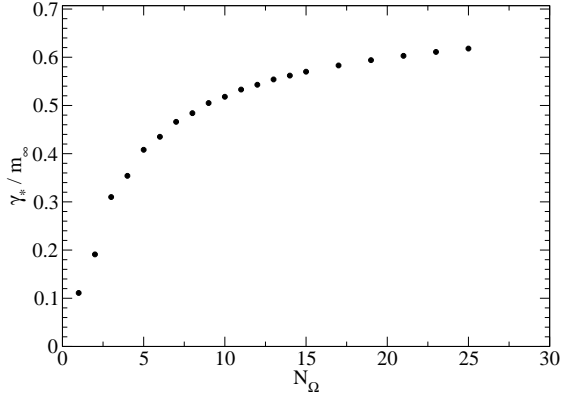}
\caption{%
    \label{fig:gamma_vs_n}
    The maximum growth rate $\gamma_*$ (in units of $m_\infty$)
    plotted vs.\ $N_\Omega$.
    The top of the graph represents the $N_\Omega \to \infty$ limit of
    $\gamma_* \to m_\infty/\sqrt{2}$ \cite{ALM}.
    }
\end{figure}

Table \ref{tab:fguy} summarizes basic properties of these
distributions for various values of $N_\Omega$.
Increasing anisotropy is signaled by decreasing
$(v_z)_{\rm rms} = \langle v_z^2\rangle^{1/2}$ and
increasing $\kmax/m_\infty$.
For graphical comparison, Fig.\ \ref{fig:gamma_vs_k} shows,
for different values of $N_\Omega$,
the perturbative growth rates of unstable modes as a function
of wavenumber $k$ in the case that $\k$ points exactly along
the beam direction.  For each distribution, $\kmax$ denotes the
largest unstable momentum, $\gamma_*$ is the largest growth rate,
and $k_*$ is the corresponding momentum.
Figs.\ \ref{fig:vz_vs_n} and \ref{fig:gamma_vs_n} show
$\kmax$, $1/(v_z)_{\rm rms}$, and $\gamma_*$ vs.\ $N_\Omega$.
For comparison, the moderately anisotropic distribution previously
simulated in Refs.\ \cite{linear1,linear2} is roughly comparable
to our $N_\Omega=3$ distribution,
and the most extremely anisotropic distribution simulated
in Ref.\ \cite{BodekerRummukainen} is roughly comparable to
our $N_\Omega=15$ distribution.%
\footnote{
   Specifically, the distribution used in Refs.\ \cite{linear1,linear2}
   has $(v_z)_{\rm rms} = 0.312$, and the
   $L_{\rm asym}=28$ distribution of Ref.\ \cite{BodekerRummukainen}
   has $(v_z)_{\rm rms} = \eta/\sqrt3 = 0.0864$.
   Also, the notation $m_l^2/m_0^2$ of
   Ref.\ \cite{BodekerRummukainen} is equivalent
   to our notation $\Omega_l$.
}

\begin {table}
\setlength{\tabcolsep}{10pt}
\begin{tabular}{|c|rrrr|}
\hline
$N_\Omega$ & $(v_z)_{\rm rms}$  & $k_{\rm max}/m_\infty$
  & $k_*/m_\infty$ & $\gamma_*/m_\infty$ \\[5pt]
\hline
 1 & 0.4472 &  1     & 0.500 & 0.111  \\
 2 & 0.3780 &  1.414 & 0.649 & 0.191  \\
 3 & 0.2852 &  2.155 & 0.875 & 0.310  \\
 4 & 0.2506 &  2.542 & 0.979 & 0.354  \\
 5 & 0.2093 &  3.149 & 1.130 & 0.408  \\
 6 & 0.1887 &  3.542 & 1.221 & 0.435  \\
 7 & 0.1653 &  4.099 & 1.342 & 0.466  \\
 8 & 0.1516 &  4.500 & 1.424 & 0.484  \\
 9 & 0.1366 &  5.031 & 1.527 & 0.505  \\
10 & 0.1269 &  5.438 & 1.602 & 0.518  \\
11 & 0.1163 &  5.954 & 1.693 & 0.533  \\
12 & 0.1091 &  6.365 & 1.763 & 0.543  \\
13 & 0.1013 &  6.870 & 1.846 & 0.554  \\
14 & 0.0957 &  7.285 & 1.911 & 0.562  \\
15 & 0.0897 &  7.783 & 1.987 & 0.570  \\
\hline
17 & 0.0805 &  8.693 & 2.119 & 0.583  \\
19 & 0.0731 &  9.601 & 2.245 & 0.594  \\
21 & 0.0668 & 10.508 & 2.361 & 0.603  \\
23 & 0.0616 & 11.414 & 2.475 & 0.611  \\
25 & 0.0571 & 12.319 & 2.583 & 0.618  \\
\hline
\end{tabular}
\caption{
  \label {tab:fguy}
  For each of the hard particle distributions designated by $N_\Omega$,
  the quantity $(v_z)_{\rm rms} = \langle v_z^2\rangle^{1/2}$
  measures the narrowness of the velocity distribution about the transverse
  plane.
  $\kmax$ is the maximum unstable wavenumber.
  $\gamma_*$ is the largest perturbative growth rate of the field
  modes $\A(\k)$ and corresponds
  to wavenumber $k_*$.
  The corresponding perturbative growth rate of magnetic energy is $2\gamma_*$.
}
\end {table}

\subsection {A reduced set of \boldmath$Y_{\ell m}$'s}

The expansion of extremely anisotropic distributions $\Omega(v_z)$
in spherical harmonics $Y_{\ell m}(\v)$
requires large $\ell$ values but, due to
the axial symmetry of the distribution, only the $m$ value
$m=0$.  The dynamics (\ref{eq:W}) of the fluctuations
$W(\v,\x,t)$ in the distribution, however, will create $W_{lm}$'s
with non-zero values of $m$.  We might hope
that only small $m$ values turn out to be significant.
Though there will be a lot of rapid variation in how $W(\v,\x,t)$
depends on $v_z$, because we are studying the case of extreme
anisotropy, there might
be relatively smooth dependence on $(v_x,v_y)$.
We will verify this picture below using simulation data, and also
give some qualitative arguments why one might expect it.
We can take advantage of this smooth dependence by
placing an upper bound $|m| \le \mmax$ on the range of $m$ we
include in our simulations,
so that
the expansion (\ref{eq:Wx0}) of $W$ becomes
\begin {equation}
   W(\x,\v,t) =
   \sum_{\ell=0}^{\lmax}
   \sum_{\begin{subarray}{c} |m|\le\ell \\ |m|\le\mmax \end{subarray}}
     W_{\ell m}(\x,t) \, \hat Y_{\ell m}(\v) ,
\label {eq:Wx}
\end {equation}
For unrestricted $m$'s, the total number of $W_{\ell m}$ at each lattice
site (and so the resources required for the simulations) would grow
quadratically with $\lmax$.
For a fixed bound $|m| \le \mmax$, however, they only grow linearly for
large $\lmax$, making simulations of extreme anisotropy practical.

\begin{figure}[t]
\includegraphics[scale=0.50]{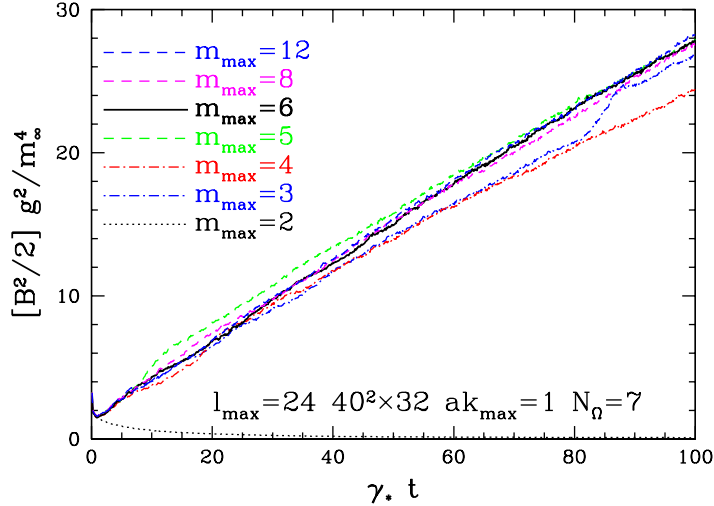}
\caption{%
    \label{fig:mmax_example}
    Linear growth of the total magnetic energy with time for several
    different values of $\mmax$.  The parameters are the same as our
    canonical $N_\Omega=7$ simulation except the box size is $40^2\times
    32$ and the squeeze factor is only 2.
}
\end{figure}

Fig.\ \ref{fig:mmax_example}
shows an example of linear growth of total magnetic
field energy for $N_\Omega=7$ simulations with
$\lmax=24$ and several different values of $\mmax$.
As can be seen, $\mmax=6$ is large enough to reproduce
the correct ($\mmax \to \infty$) slope,
and this is the value of $\mmax$ we will use in our
simulations.

In previous work on simulations for moderate anisotropy \cite{linear2},
we found that the systematic errors arising from a finite cut off
$\lmax$ on $\ell$ could be reduced by damping the dynamics of modes with
$\ell$ near the cut-off.
We have slightly improved this method and extended it to apply also to
the new cut-off $\mmax$ on $m$.
Details are given in Appendix \ref{app:damp}.
Such damping has been used in all the simulations reported in this paper.

The real test of the viability of using relatively small $\mmax$
cut-offs comes from simulations, such as Fig.\ \ref{fig:mmax_example}.
However, one can get some rough idea of why it can work
by considering perturbative formulas for some of the important
features of unstable modes and the resulting cascade of plasmons.
If we treat the gauge field perturbatively in the $W$ equation
(\ref{eq:W}),
replacing $D_\mu$ by $\partial_\mu$, we can Fourier transform
from $(\x,t)$ to $(\k,\omega)$ and then solve for $W$:
\begin {equation}
  W(\v,\k,\omega) =
   i m_\infty^2 \; (\omega-\v\cdot\k)^{-1}
   \left[ \E(\k,\omega) \cdot (2\v-\grad_\v)
     + \B(\k,\omega) \cdot ( \v \times \grad_\v ) \right] \Omega(v_z) \, .
\end {equation}
Together, the factors to the right of the $(\omega-\v\cdot\k)^{-1}$
in this formula only generate $\v_\perp$ dependence with $|m|\le 1$.
All higher $m$ components in the result for $W$ are generated by the
factor
\begin {equation}
   (\omega - \v\cdot\k)^{-1}
   = (\omega - \v_\perp\cdot\k_\perp - v_z k_z)^{-1} .
\label {eq:factor}
\end {equation}

Now consider the dominant unstable mode.
As mentioned earlier, this mode has $\k$ along the $z$ axis
(for the type of anisotropy we consider), and so $\k_\perp = 0$.
Then the factor (\ref{eq:factor}) has no $\v_\perp$ dependence, and
so the $W$ field which describes the dominant instability
involves only $|m| \le 1$.

As another example, consider the dispersion relation of transverse plasmons.
A standard method for deriving the dispersion relation is to
insert the result for $W$ into the Yang-Mills equation (\ref{eq:YangMills}),
which generates the hard-loop self-energy correction to the
vacuum relation $\omega^2 = k^2$.
How much will we disturb this dispersion relation if we throw
away modes of $W$ with $m > \mmax$?
For high momentum plasmons ($k \gg m_\infty$), such as those that
dominate the cascade at late times, the effect is tiny simply because
all medium effects to the dispersion relation are tiny in this limit.
For very low momentum plasmons ($k \ll \omega \sim m_\infty$),
we can ignore the $\v\cdot\k$ altogether in (\ref{eq:factor}), and
then the $W$ field will again have only $|m| \le 1$ components.
It is only for intermediate momentum plasmons ($k \sim m_\infty$)
that finite $\mmax$ does violence to the plasmon dispersion
relation.  However, (\ref{eq:factor}) is a fairly
smooth function of $\v_\perp$
in this regime because the denominator never gets close
to zero for $k \sim m_\infty$ plasmons
($\omega$ and $\omega-k$ are both of order $m_\infty$),
and so (\ref{eq:factor}) and therefore $W$ can be reasonably
approximated by a superposition of relatively low $m$'s.

\begin{figure}[t]
\includegraphics[scale=0.50]{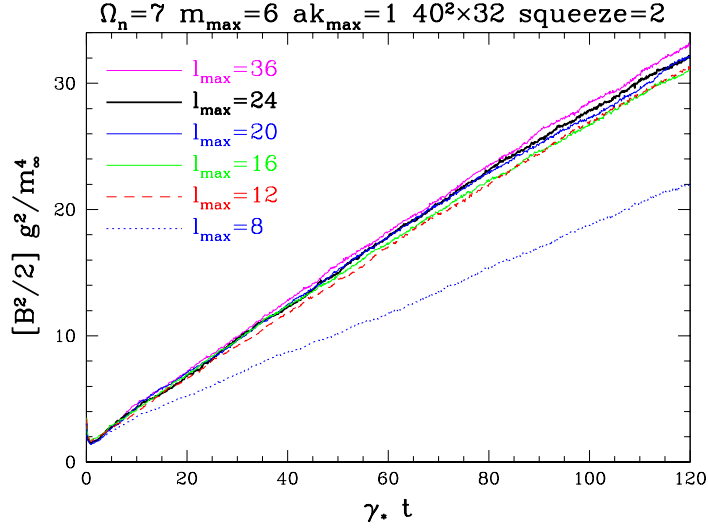}
\caption{%
    \label{fig:lmax_n7}
    Linear growth of the total magnetic energy with time for several
    different values of $\lmax$, and fixed $\mmax=6$, for
    the hard particle distribution $N_\Omega = 7$.
}
\end{figure}

Finally, we should check that we have chosen large enough values
of $\lmax$ in our simulations.  In general, we find that
$\lmax \sim 3 N_\Omega$ is quite adequate.
As an example, Fig.\ \ref{fig:lmax_n7} shows 
the $\lmax$ dependence of
$N_\Omega=7$ simulations for fixed
$\mmax=6$.  Our standard simulation choice for $\lmax$
is 24 for $N_\Omega = 7$.
See Table \ref{tab:params} for our default simulation parameters
in other cases.

\begin {table}[t]
\setlength{\tabcolsep}{10pt}
\begin{tabular}{|c|llll|}
\hline
$N_\Omega$ & $a\kmax$  & $\lmax$ & squeeze $s$ & volume \\[5pt]
\hline
 3 & 0.6 & 24 & 1.5 & $64^2\times32$ \\
 4 & 0.6 & 24 & 2   & $64^2\times32$ \\
 5 & 0.8 & 24 & 2.5 & $64^2\times32$ \\
 6 & 0.8 & 24 & 3   & $64^2\times32$ \\
 7 & 1.0 & 24 & 3.5 & $64^2\times32$ \\
 8 & 1.0 & 32 & 3.5 & $64^2\times32$ \\
 9 & 1.0 & 32 & 3.5 & $64^2\times32$ \\
11 & 1.0 & 40 & 3.5 & $64^2\times32$ \\
13 & 1.0 & 48 & 3.5 & $64^2\times32$ \\
15 & 1.0 & 56 & 3.5 & $64^2\times32$ \\
25 & 1.2 & 80 & 3   & $64^2\times28$ \\ \hline
 5 & 0.6 & 24 & 2.5 & $64^2\times32$ \\
 6 & 0.6 & 24 & 3.0 & $64^2\times32$ \\
 7 & 0.6 & 24 & 3.5 & $64^2\times32$ \\
 7 & 0.8 & 24 & 3.0 & $64^2\times32$ \\
\hline
\end{tabular}
\caption{
  \label {tab:params}
  The default parameters for our simulations and their initialization,
  as a function of $N_\Omega$.
  The corresponding values of $\kmax$ are given in Table \ref{tab:fguy}.
  Other default parameters include
  $\mmax = 6$, initial temperature $T=\kmax/g^2$,
  and initial smearing wavenumber $k_{\rm smear} = \kmax$.
  The simulations below the horizontal line correspond to the crosses
  in Fig.\ \ref{fig:dEdt}.
}
\end {table}


\subsection {Initial conditions}
\label {sec:strong}

Following Ref.\ \cite{linear2}, we use strong, non-perturbative initial
conditions for the magnetic field $\B$, so that the system starts
linear energy growth behavior as quickly as possible.
The electric and $W$ fields are, for simplicity, initialized to zero.
In order to see the linear energy growth associated with cascade
development as early as possible, it is advantageous to choose initial
conditions which do not significantly populate modes with large
wavenumber.

In the moderate anisotropy simulations of Ref.\ \cite{linear2},
the initial magnetic field was constructed by taking a thermal
initial state with temperature $T = 2m_\infty/g^2$ and then
performing gauge-invariant smearing (sometimes called cooling) of
the configuration to eliminate wavenumbers $k \gg m_\infty$.
In perturbative language, this cooling corresponds to replacing
the initial thermal field $\A_{\rm therm}$ by
\begin {equation}
   \A(\k) = \A_{\rm therm}(\k) \exp(-k^2/k_{\rm smear}^2),
\end {equation}
where
$\tau = 1/k_{\rm smear}^2$ is the
smearing parameter.  In Ref.\ \cite{linear2}, we chose
$k_{\rm smear} = 2m_\infty$.

Here, we follow a similar procedure, but the unstable modes that we
want to initially populate are generally more extremely anisotropic, having
$(k_\perp, k_z) \sim (m_\infty, \kmax)$ with $\kmax \gg m_\infty$.
We have found that it helps to arrange a related anisotropy of
our initial fields by squeezing the initial distribution in
the $z$ direction.  In perturbative language, our initial choice
corresponds to
\begin {equation}
  (\A_\perp,A_z)[\k] = (\A_{\perp,{\rm therm}}/s,A_{z,{\rm therm}})
    [s\k_\perp,k_z]\exp\Big(-(s^2 k_\perp^2+k_z^2)/k_{\rm smear}^2\Big),
\end {equation}
where $s$ is the squeezing factor.%
\footnote{
  Our technical procedure is to choose the initial
  magnetic field by the usual procedure but pretending that the
  lattice is asymmetric with lattice spacing $a_\perp = a/s$ 
  in the transverse directions,
  compared to $a$ along the $z$ axis.  We then re-interpret
  the resulting initial condition as living on the symmetric lattice
  ($a_z = a_\perp$) used in our simulations.
}
In our simulations, we have generally chosen $T=\kmax/g^2$,
$k_{\rm smear}=\kmax$, and $s$ between $1.5$
and $3.5$ depending on the amount of anisotropy.
See Table \ref{tab:params} for our default simulation parameters.


\subsection {Lattice spacing and volume}

It is important to check that the lattice volume is large enough to be
in the infinite volume limit and the spacing is small enough to be in
the continuum limit; otherwise the lattice calculation is not simulating
the desired continuum physics.  It would be prohibitive to check this at
every lattice spacing, so we have ``spot checked'' this at a few levels
of anisotropy, with the most thorough study at $N_\Omega=7$ and
$N_\Omega = 15$.

For highly anisotropic lattices, the physical scales possibly relevant
to out problem parametrically span a range from $m_\infty$ to $\kmax$.
One might worry that, as particle distributions are taken more and more
anisotropic, it becomes harder and harder to span these scales with a
computationally practical lattice.  Naively, to be perfectly safe, we
would like physical lattice dimensions $L \gg 2\pi/m_\infty$ and lattice
spacings $a \ll 2\pi/\kmax$.  In this section, we'll see how well
we do with lattices of practical size.


\subsubsection{Physical volume}

\begin{figure}[t]
\includegraphics[scale=0.50]{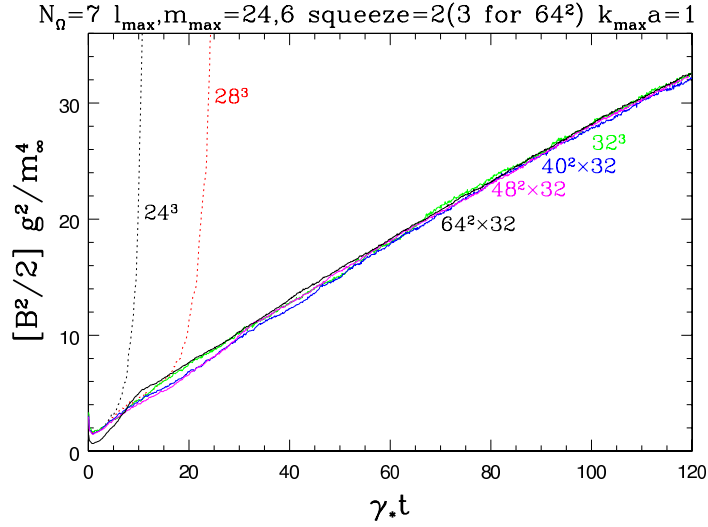}
\caption{%
    \label{fig:vol_n7}
    Linear growth of the total magnetic energy with time for several
    different physical volumes, at fixed lattice spacing, for
    the hard particle distribution $N_\Omega = 7$.
}
\end{figure}

\begin{figure}[t]
\includegraphics[scale=0.50]{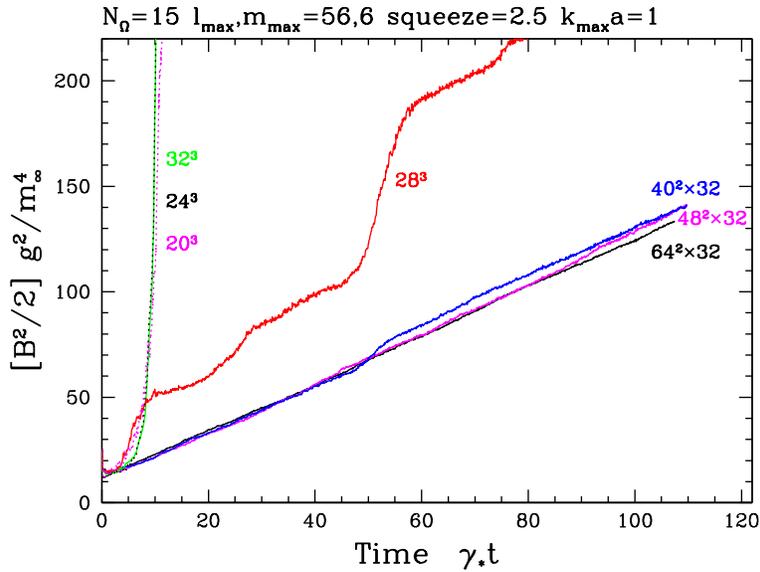}
\caption{%
    \label{fig:vol_n15}
    As Fig.\ \ref{fig:vol_n7} but for the more anisotropic distribution
    $N_\Omega = 15$.
}
\end{figure}

Figs.\ \ref{fig:vol_n7} and \ref{fig:vol_n15} show
the volume dependence, at fixed lattice spacing, of the evolution of
magnetic energy with time.  The first figure is for the
$N_\Omega{=}7$ hard particle distribution.  The second figure is
for $N_\Omega{=}15$, the second most anisotropic distribution included
in our results of Fig.\ \ref{fig:dEdt}.
Our default lattice size of $64^2\times32$ corresponds to approximately
$(15.6/m_\infty)^2\times(7.8/m_\infty)$ for
$N_\Omega=7$ and $(8.2/m_\infty)^2\times(4.1/m_\infty)$ for
$N_\Omega=15$.
For small volumes, the
simulations produced exponential rather than linear growth.%
\footnote{
   These are small volumes with periodic boundary conditions.
   One should not expect this small-volume exponential growth behavior
   for a comparably small volume of hard particles in infinite space,
   surrounded by vacuum.  In that case the hard particles
   would escape the small volume
   within the time scale characteristic of the instability growth.
}
But linear growth appears at large enough volume, and a comparison of
the large volume curves suggests that our default lattice size
of $64^2\times32$ is adequate, even for our highly anisotropic
distributions.

\begin{figure}[t]
\includegraphics[scale=0.50]{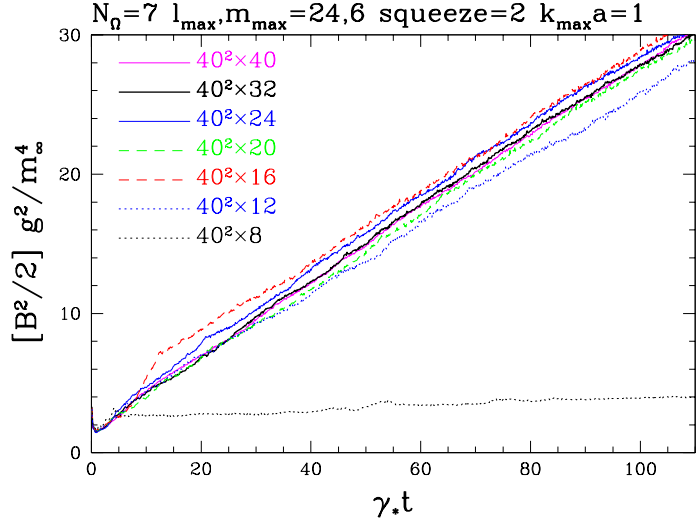}
\caption{%
    \label{fig:zval}
    Linear growth of the total magnetic energy with time on an
    $L_\perp^2 \times L_z$ lattice for several different choices
    of $L_z$ (at fixed lattice spacing),
    for
    the hard particle distribution $N_\Omega = 7$.
}
\end{figure}

In order to be able to run our simulations on desktop computers, we
by default took the physical lattice size $L_z$
in the $z$ direction
to be half that in the $x$ and $y$ directions.
This choice is motivated by the fact that, in the highly anisotropic case,
unstable modes have parametrically smaller wavelength in the $z$ direction
($\sim 1/\kmax$) than in the perpendicular directions ($\sim 1/m_\infty$).
Of course, that doesn't exclude the possibility that stable modes
with size $k_x \sim k_y \sim k_z \sim 1/m_\infty$ might be important
in the development of linear growth, and so we should investigate
the matter with simulations. 
Fig.\ \ref{fig:zval}
isolates the effect of varying $L_z$ while holding the $L_x$ and $L_y$
fixed.  (Note that $L_x=L_y$ is smaller here than in
Fig.\ \ref{fig:vol_n7}.)
An $L_z$ that is half of $L_x=L_y$ appears adequate for reproducing the
large-$L_z$ linear slope.

\subsubsection {Lattice spacing}

\begin{figure}[t]
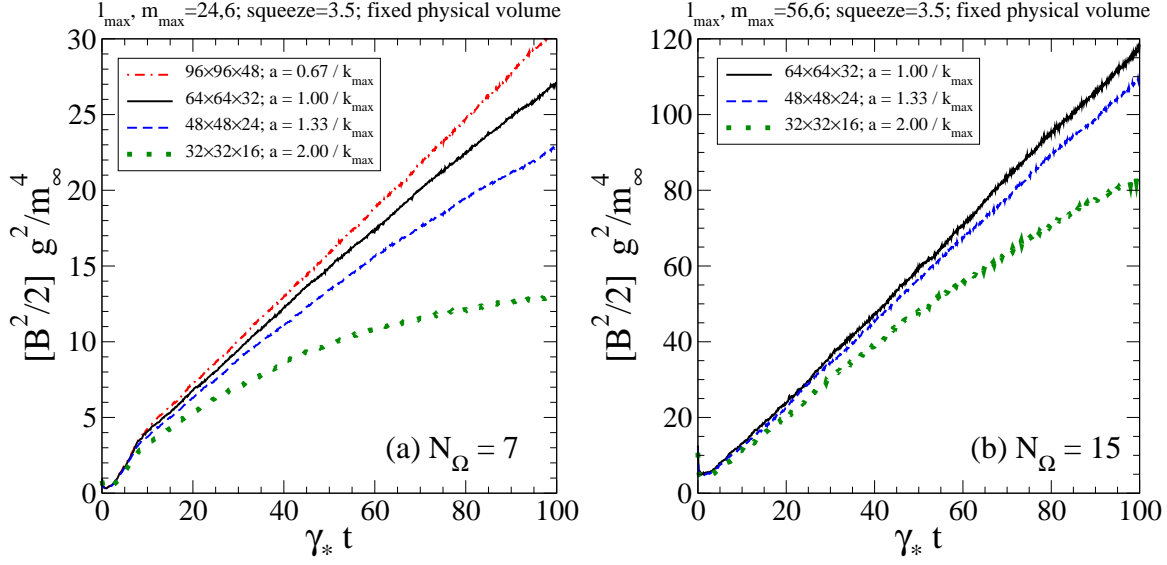

\includegraphics[scale=0.40]{spacing2_n7.eps}
\includegraphics[scale=0.40]{spacing2_n15.eps}
\caption{%
    \label{fig:spacing}
    Linear growth of the total magnetic energy with time for several
    different lattice spacings, at fixed physical volume
    $L_x{\times}L_y{\times}L_z
    = (64/\kmax){\times}(64/\kmax){\times}(32/\kmax)$,
    for the hard particle distributions (a) $N_\Omega = 7$
    and (b) $N_\Omega = 15$.
}
\end{figure}

Fig.\ \ref{fig:spacing} shows
how our simulations depend on lattice spacing for fixed physical
volume, for the distributions $N_\Omega=7$ and $N_\Omega=15$.
In order to isolate the effect of lattice spacing, we
have used the same initial conditions for all these simulations.
More precisely, we generated initial conditions for
the finest lattice ($96{\times}96{\times}48$),
and then we used blocking to generate similar
initial conditions for the coarser lattices.%
\footnote{
  For instance, to block by a factor of 2 in every direction, one could
  replace appropriate pairs $U_1$ and $U_2$ of consecutive links by
  a single link $U_1 U_2$.  In practice, we use the slightly improved
  method of
  averaging this with the four ``staples'' that move one link
  transversely, then two links in the direction of interest, and
  then back again transversely.  We use a similar method for blocking
  by 3 and then iterate as necessary to get the various lattice
  sizes used.
  (We did not simulate the evolution of
  $96{\times}96{\times}48$ for $N_\Omega=15$ because
  of memory limitations.)
}

At all but the finest lattice spacing in Fig.\ \ref{fig:spacing}a,
one can see
some curvature to the late-time ``linear'' growth behavior.
This curvature is a lattice artifact, but it means that we
need a procedure for extracting a single ``slope'' from
such simulations, in order to present results such as
Fig.\ \ref{fig:dEdt}.
Note that the coarsest lattice spacing results in
Fig.\ \ref{fig:spacing} look like
sections of tanh curves, after an initial transient.
Inspired by this observation, we have chosen to fit each
of our energy curves to the form
\begin {equation}
   \half B^2(t) = \sigma \, t_1 \tanh\left( \frac{t-t_0}{t_1} \right)
\label {eq:tanh_fit}
\end {equation}
for $\gamma_* t > 10$.  The parameters of the fit are
$s$, $t_0$, and $t_1$.  We take the slope $\sigma$ (the slope of the
tanh at zero argument) to be our result for $d{\cal E}^B_{\rm tot}/dt$.
The curving of the tanh is controlled by
$t_1$, and $t_1$ should go to infinity as we approach the continuum limit.

\begin{figure}[t]
\includegraphics[scale=0.40]{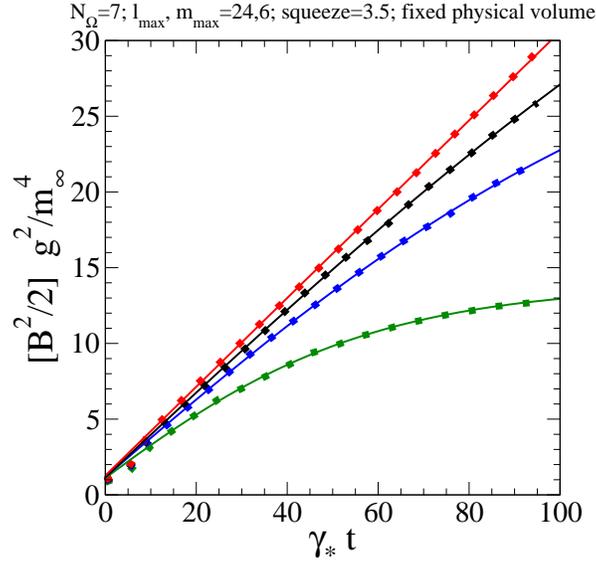}
\caption{%
    \label{fig:fit}
    The curves of Fig.\ \ref{fig:spacing}a, all shown here as dotted lines,
    superposed with solid lines corresponding to the fits of
    Eq.\ (\ref{eq:tanh_fit}).
}
\end{figure}

\begin{figure}[t]
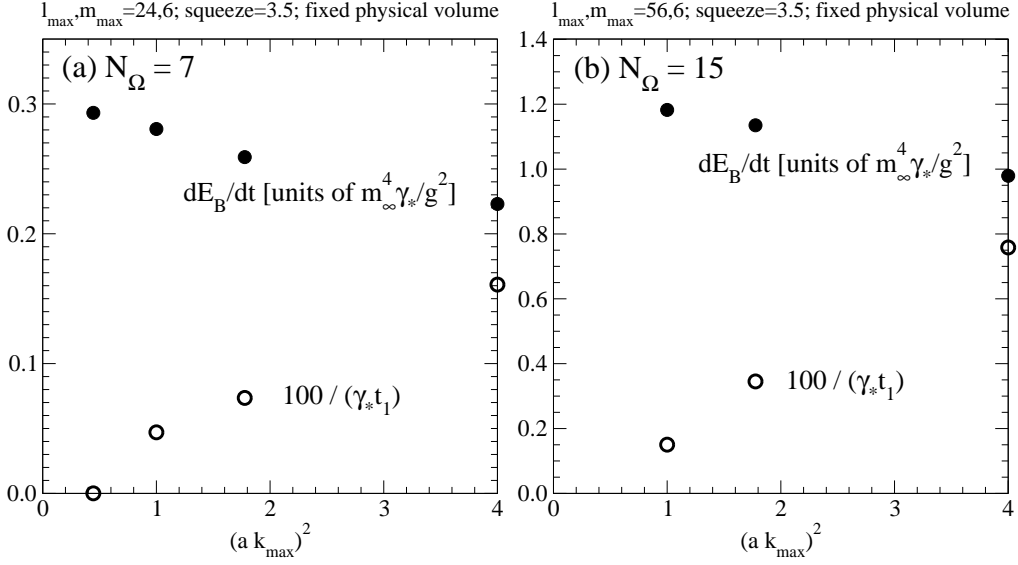

\includegraphics[scale=0.40]{extrapolate3_n7.eps}
\includegraphics[scale=0.40]{extrapolate3_n15.eps}
\caption{%
    \label{fig:extrapolate}
    Fit parameters for the (a) $N_\Omega = 7$ and (b) $N_\Omega = 15$
    simulations of
    Fig.\ \ref{fig:spacing} as a function of lattice spacing squared.
    Solid circles are the slope $\sigma$ (and so $d{\cal E}^B_{\rm tot}/dt$)
    in units of $m_\infty^4 \gamma_*/g^2$, and open circles are
    $1/t_1$ in units of $(100/\gamma_*)^{-1}$.
}
\end{figure}

Fig.\ \ref{fig:fit} shows the tanh fits for the simulations of
Fig.\ \ref{fig:spacing}a: the fits work extremely well.
The solid circles in Fig.\ \ref{fig:extrapolate}
show how the fit of the slope $\sigma$
depends on the lattice spacing.  The $x$ axis is chosen to be the square
of the lattice spacing because the discretization errors in our lattice
implementation first arise at this order.  Extrapolating by eye to the
continuum limit, we estimate that our default lattice spacing
of $a \kmax = 1$ for these distributions has lattice spacing
errors no larger than roughly 10\%.
In contrast, the open circles in Figs.\ \ref{fig:extrapolate}
show the behavior of $1/t_1$, which is a lattice
artifact and approaches zero in the continuum limit
(corresponding to purely linear growth).

\section {Conclusion}

The goal of this paper has been to understand how, in the weak
coupling limit, the late-time
behavior of Weibel instabilities scales with hard particle anisotropy.
We can use the smallness of $\theta \equiv v_z$, characterizing the angular
distribution of hard particles, as a measure of anisotropy.
Through simulations, we have examined the slope
$d{\cal E}_{\rm tot}^B/dt$ of the late-time linear growth in
the total magnetic energy of soft gauge fields
and found that the scaling of this slope is consistent
with $\theta^{-2}$ and not consistent with $\theta^{-1}$ or $\theta^{-3}$.
If we accept the simple model outlined in section
\ref{sec:measure}
of the physics behind
$d{\cal E}_{\rm tot}^B/dt$, this result implies that
the limiting magnetic field strength $B_*$ of Weibel unstable
modes scales with anisotropy as $\theta^{-1}$ and is of order
\begin {equation}
   B_* \sim \frac{m_\infty^4}{\theta g} \,.
\end {equation}
Of course, it would be better not to rely on such indirect arguments.
A goal for future work should be to check the consistency of
this conclusion with alternative measurements.


\begin{acknowledgments}

We thank Michael Strickland, Kari Rummukainen, Dietrich B\"odeker,
and Berndt M\"uller
for useful conversations.
This work was supported, in part, by the U.S. Department
of Energy under Grant No.~DE-FG02-97ER41027,
by the National Sciences and Engineering
Research Council of Canada, and by le Fonds Nature et Technologies du
Qu\'ebec.

\end{acknowledgments}

\appendix

\section{The Nielsen-Olesen limit}
\label {app:NO}

At one time, it was conjectured that
exponential instability growth would continue beyond
the point where non-abelian interactions became important because
the fields would dynamically align themselves into a
commuting set of color directions \cite{AL}.
The fields would then be effectively abelian and could continue
growing, just like the purely abelian case shown by the
dashed line in Fig.\ \ref{fig:linear1}.
This conjecture seemed borne out by early
simulations in one spatial dimension \cite{RRS},
such as shown by the dotted line in that figure.
At the time,
Berndt M\"uller \cite{berndt} predicted that three-dimensional instability
growth would eventually have to stop because, even if the gauge fields
did abelianize,
Nielsen-Olesen instabilities \cite{NielsenOlesen}
would eventually destroy nearly-abelian
configurations as the fields continue to grow.

In this appendix, we will discuss the largest field strength allowed for
nearly-abelian Weibel unstable modes before Nielsen-Olesen instabilities
appear.  A brief review of Nielsen-Olesen instabilities in the context
of Weibel instabilities for moderate anisotropy can be found in Ref.\
\cite{ArnoldLeang}.  Here, we generalize to the case of extreme
anisotropy.

First imagine a situation where there was a large, constant,
homogeneous magnetic field $B_0$ that lies within an abelian
subgroup of the non-abelian gauge group.
In our application, this could represent a large magnetic field
that was created by the Weibel instability and that might have
abelianized due to non-linear dynamics, and that we are looking
at this field on small enough time and distance scales that we can
treat it as constant.
For the sake of definiteness, consider SU(2) gauge theory and
a background magnetic field
\begin {equation}
   B_i^a(\x,t) = \B_0 \, \delta^{a3} ,
\end {equation}
where $a$ is the adjoint color index.
Now one can investigate the dispersion relation of fluctuations about
this background field, including fluctuations involving other,
non-commuting color directions.  Ignoring hard particle effects,
the result is \cite{NielsenOlesen}
\begin {equation}
   \omega^2 = q_\parallel^2 + (n+\half) 2 |Q| g B_0 - 2\ms Q g B_0 ,
\label {eq:NOdisp}
\end {equation}
where $q_\parallel$ is the component of momentum parallel to
$\B_0$; $n$ is the Landau orbit quantum number for circular
motion in the plane transverse to $\B_0$; $\ms = 0$ or $\pm1$ is the
component, in the direction of $\B_0$, of the spin of a
gauge excitation; and $Q=0$ or $\pm1$ is the charge of that excitation
under the color generator $T^3$.  The last term in (\ref{eq:NOdisp})
represents the interaction energy of a gauge particle's magnetic
moment with the magnetic field.  If we look at the lowest
Landau orbits ($n=0$) and the sector
$Q = \ms = \pm1$, we get
\begin {equation}
   \omega^2 = q_\parallel^2 - g B_0 .
\label {eq:NOomega}
\end {equation}
$\omega^2$ is then negative for $q_\parallel \le \sqrt{g B_0}$.
This is the Nielsen-Olesen instability.

In our application, the magnetic fields are not homogeneous.
In order to make that approximation, the radius
associated with the lowest Landau orbit should be small
enough to fit in a region
of roughly constant magnetic field.
This radius is $R \sim 1/\sqrt{g B_0}$.
The dominant Weibel-unstable modes
have $B$ roughly orthogonal to the $z$ axis, and wavenumber of order
$(k_\perp, k_z) \sim (m_\infty, \kmax)$.
So, to produce the Nielsen-Olesen instability, we need both
$R \ll 1/k_\perp$ and
$R \ll 1/k_z$.
In the case of extreme anisotropy, the latter is the stricter
constraint, equivalent to
\begin {equation}
  \frac{1}{\sqrt{gB_0}} \ll \frac{1}{\kmax} \sim
  \frac{\iso}{m_\infty}
  \,,
\end {equation}
which requires
\begin {equation}
   B_0 \gg \frac{m_\infty^2}{\iso^2 g} \,.
\label {eq:NOconstraint}
\end {equation}
As the magnetic field grows, Nielsen-Olesen
instabilities will then first appear
for
\begin {equation}
   B_0 \sim \frac{m_\infty^2}{\iso^2 g} \,.
\label {eq:NOlimit}
\end {equation}
This corresponds to the case
$\expon=2$ in (\ref{eq:Bn}).

There are a few approximations left to check.
First, the Nielsen-Olesen analysis assumed that $B_0$ was
constant in time.  The typical Nielsen-Olesen instability
growth times generated by (\ref{eq:NOomega}) will be
of order $(-\omega^2)^{-1/2} \sim 1/\sqrt{g B}$.  This
will be small compared to the (abelianized) Weibel instability growth
time $\gamma_*^{-1} \sim m_\infty^{-1}$ when (\ref{eq:NOconstraint}) is
satisfied.  Secondly, we have ignored hard particle effects
throughout.
In general, the magnitude of the soft self-energy
$\Pi$ due to hard particles can be as large as
order $\kmax^2 \sim m_\infty^2/\iso^2$, depending on direction
and $\omega/k$.
But this is small enough that $\Pi$ will be a small correction
to $g B_0$ in the dispersion relation
(\ref{eq:NOomega}) when (\ref{eq:NOconstraint}) is satisfied.

The results of this paper suggest that $\expon=1$ rather than
$\expon=2$.  It may well be that Nielsen-Olesen effects limit the Weibel
instability growth of nearly abelian fields.
The results of this paper simply suggest that, if one starts
with large amplitude, nonperturbative,
{\it non-abelian}\/ initial conditions, then
generic non-perturbative interactions are sufficient to
stop growth earlier, at lower field strength than (\ref{eq:NOlimit}).


\section{The coefficients \boldmath$\Omega_\ell$}
\label {app:Omega}

In principle, anyone who wanted to know the specific values of
$\Omega_{l}$ for our distributions could reproduce them from the
procedure outlined in the text.  However, we found that avoiding
numerical round-off errors in determining $\Omega_{l}$ for large
$N_\Omega$ required some care.  So we will explicitly give our
distributions here.

\medskip



$N_\Omega=1$:
$\Omega_0= 1$,
$\Omega_2= - 1/\sqrt{5}$

$N_\Omega=2$:
$\Omega_0= 1$,
$\Omega_2= -2\sqrt{5}/7$,
$\Omega_4= 1/7$

$N_\Omega=3$:
$\alpha_{1} = 0.585310$;
$\Omega_0= 1$,
$\Omega_{2}=    -0.845154$,
$\Omega_{4}=     0.474960$,
$\Omega_{6}=    -0.148398$

$N_\Omega=4$:
$\alpha_{1} = 0.482673$;
$\Omega_0= 1$,
$\Omega_{2}=    -0.907458$,
$\Omega_{4}=     0.589250$,
$\Omega_{6}=    -0.277350$,
$\Omega_{8}=     0.063395$

$N_\Omega=5$:
$\alpha_{1} = 0.350109$,
$\alpha_{2} = 0.759931$;
$\Omega_0= 1$,
$\Omega_{2}=    -0.971104$,
$\Omega_{4}=     0.730670$,
$\Omega_{6}=    -0.449643$,
$\Omega_{8}=     0.216519$,
$\Omega_{10}=    -0.063736$

$N_\Omega=6$:
$\alpha_{1} = 0.292253$,
$\alpha_{2} = 0.672147$;
$\Omega_0= 1$,
$\Omega_{2}=    -0.998631$,
$\Omega_{4}=     0.794155$,
$\Omega_{6}=    -0.544163$,
$\Omega_{8}=     0.311154$,
$\Omega_{10}=    -0.138336$,
$\Omega_{12}=     0.032712$

$N_\Omega=7$:
$\alpha_{1} = 0.228412$,
$\alpha_{2} = 0.545787$,
$\alpha_{3} = 0.845543$;
$\Omega_0= 1$,
$\Omega_{2}=    -1.026410$,
$\Omega_{4}=     0.864186$,
$\Omega_{6}=    -0.650680$,
$\Omega_{8}=     0.436901$,
$\Omega_{10}=    -0.249727$,
$\Omega_{12}=     0.113554$,
$\Omega_{14}=    -0.032709$

$N_\Omega=8$:
$\alpha_{1} = 0.194863$,
$\alpha_{2} = 0.478948$,
$\alpha_{3} = 0.776881$;
$\Omega_0= 1$,
$\Omega_{2}=    -1.040916$,
$\Omega_{4}=     0.901356$,
$\Omega_{6}=    -0.712570$,
$\Omega_{8}=     0.512936$,
$\Omega_{10}=    -0.329365$,
$\Omega_{12}=     0.179614$,
$\Omega_{14}=    -0.077926$,
$\Omega_{16}=     0.018926$

$N_\Omega=9$:
$\alpha_{1} = 0.159625$,
$\alpha_{2} = 0.400532$,
$\alpha_{3} = 0.671219$,
$\alpha_{4} = 0.892835$;
$\Omega_0= 1$,
$\Omega_{2}=    -1.055491$,
$\Omega_{4}=     0.940673$,
$\Omega_{6}=    -0.778554$,
$\Omega_{8}=     0.600618$,
$\Omega_{10}=    -0.424975$,
$\Omega_{12}=     0.271564$,
$\Omega_{14}=    -0.149879$,
$\Omega_{16}=     0.066253$,
$\Omega_{18}=    -0.018902$,

$N_\Omega=10$:
$\alpha_{1} = 0.138840$,
$\alpha_{2} = 0.353981$,
$\alpha_{3} = 0.608348$,
$\alpha_{4} = 0.839259$;
$\Omega_0= 1$,
$\Omega_{2}=    -1.064055$,
$\Omega_{4}=     0.963974$,
$\Omega_{6}=    -0.819732$,
$\Omega_{8}=     0.656358$,
$\Omega_{10}=    -0.490737$,
$\Omega_{12}=     0.337910$,
$\Omega_{14}=    -0.210147$,
$\Omega_{16}=     0.111992$,
$\Omega_{18}=    -0.047971$,
$\Omega_{20}=     0.011893$

$N_\Omega=11$:
$\alpha_{1} = 0.117460$,
$\alpha_{2} = 0.303204$,
$\alpha_{3} = 0.531249$,
$\alpha_{4} = 0.753079$,
$\alpha_{5} = 0.921475$;
$\Omega_0= 1$,
$\Omega_{2}=    -1.072643$,
$\Omega_{4}=     0.988135$,
$\Omega_{6}=    -0.862564$,
$\Omega_{8}=     0.717049$,
$\Omega_{10}=    -0.563740$,
$\Omega_{12}=     0.417060$,
$\Omega_{14}=    -0.285605$,
$\Omega_{16}=     0.177636$,
$\Omega_{18}=    -0.096139$,
$\Omega_{20}=     0.041819$,
$\Omega_{22}=    -0.011875$

$N_\Omega=12$:
$\alpha_{1} = 0.103801$,
$\alpha_{2} = 0.270617$,
$\alpha_{3} = 0.481526$,
$\alpha_{4} = 0.697580$,
$\alpha_{5} = 0.879018$;
$\Omega_0= 1$,
$\Omega_{2}=    -1.078115$,
$\Omega_{4}=     1.003612$,
$\Omega_{6}=    -0.890930$,
$\Omega_{8}=     0.757648$,
$\Omega_{10}=    -0.614860$,
$\Omega_{12}=     0.473819$,
$\Omega_{14}=    -0.343760$,
$\Omega_{16}=     0.230949$,
$\Omega_{18}=    -0.141010$,
$\Omega_{20}=     0.074196$,
$\Omega_{22}=    -0.031545$,
$\Omega_{24}=     0.007947$

$N_\Omega=13$:
$\alpha_{1} = 0.089898$,
$\alpha_{2} = 0.236254$,
$\alpha_{3} = 0.425611$,
$\alpha_{4} = 0.627277$,
$\alpha_{5} = 0.808562$,
$\alpha_{6} = 0.940062$;
$\Omega_0= 1$,
$\Omega_{2}=    -1.083597$,
$\Omega_{4}=     1.019484$,
$\Omega_{6}=    -0.920055$,
$\Omega_{8}=     0.800595$,
$\Omega_{10}=    -0.669524$,
$\Omega_{12}=     0.537151$,
$\Omega_{14}=    -0.410252$,
$\Omega_{16}=     0.296157$,
$\Omega_{18}=    -0.198861$,
$\Omega_{20}=     0.121655$,
$\Omega_{22}=    -0.065052$,
$\Omega_{24}=     0.028013$,
$\Omega_{26}=    -0.007936$

$N_\Omega=14$:
$\alpha_{1} = 0.080480$,
$\alpha_{2} = 0.212901$,
$\alpha_{3} = 0.387457$,
$\alpha_{4} = 0.579149$,
$\alpha_{5} = 0.760545$,
$\alpha_{6} = 0.905790$;
$\Omega_0= 1$,
$\Omega_{2}=    -1.087305$,
$\Omega_{4}=     1.030254$,
$\Omega_{6}=    -0.940283$,
$\Omega_{8}=     0.830602$,
$\Omega_{10}=    -0.708871$,
$\Omega_{12}=     0.583371$,
$\Omega_{14}=    -0.460883$,
$\Omega_{16}=     0.347161$,
$\Omega_{18}=    -0.247263$,
$\Omega_{20}=     0.163617$,
$\Omega_{22}=    -0.098747$,
$\Omega_{24}=     0.051563$,
$\Omega_{26}=    -0.021818$,
$\Omega_{28}=     0.005568$

$N_\Omega=15$:
$\alpha_{1} = 0.070949$,
$\alpha_{2} = 0.188716$,
$\alpha_{3} = 0.346338$,
$\alpha_{4} = 0.523712$,
$\alpha_{5} = 0.698217$,
$\alpha_{6} = 0.847595$,
$\alpha_{7} = 0.952782$;
$\Omega_0= 1$,
$\Omega_{2}=    -1.091017$,
$\Omega_{4}=     1.041226$,
$\Omega_{6}=    -0.960895$,
$\Omega_{8}=     0.861825$,
$\Omega_{10}=    -0.750085$,
$\Omega_{12}=     0.633157$,
$\Omega_{14}=    -0.516209$,
$\Omega_{16}=     0.405183$,
$\Omega_{18}=    -0.303840$,
$\Omega_{20}=     0.215905$,
$\Omega_{22}=    -0.143127$,
$\Omega_{24}=     0.086609$,
$\Omega_{26}=    -0.045945$,
$\Omega_{28}=     0.019653$,
$\Omega_{30}=    -0.005561$

$N_\Omega=17$:
$\alpha_{1} = 0.057385$,
$\alpha_{2} = 0.153941$,
$\alpha_{3} = 0.286217$,
$\alpha_{4} = 0.440599$,
$\alpha_{5} = 0.601196$,
$\alpha_{6} = 0.751477$,
$\alpha_{7} = 0.875973$,
$\alpha_{8} = 0.961858$;
$\Omega_0= 1$,
$\Omega_{2}=    -1.096274$,
$\Omega_{4}=     1.056905$,
$\Omega_{6}=    -0.990855$,
$\Omega_{8}=     0.907738$,
$\Omega_{10}=    -0.812101$,
$\Omega_{12}=     0.709497$,
$\Omega_{14}=    -0.603863$,
$\Omega_{16}=     0.499893$,
$\Omega_{18}=    -0.400800$,
$\Omega_{20}=     0.310023$,
$\Omega_{22}=    -0.229608$,
$\Omega_{24}=     0.161402$,
$\Omega_{26}=    -0.106052$,
$\Omega_{28}=     0.063689$,
$\Omega_{30}=    -0.033601$,
$\Omega_{32}=     0.014305$,
$\Omega_{34}=    -0.004046$

$N_\Omega=19$:
$\alpha_{1} = 0.047353$,
$\alpha_{2} = 0.127821$,
$\alpha_{3} = 0.239919$,
$\alpha_{4} = 0.374139$,
$\alpha_{5} = 0.519102$,
$\alpha_{6} = 0.662513$,
$\alpha_{7} = 0.792211$,
$\alpha_{8} = 0.897196$,
$\alpha_{9} = 0.968556$;
$\Omega_0= 1$,
$\Omega_{2}=    -1.100133$,
$\Omega_{4}=     1.068574$,
$\Omega_{6}=    -1.013442$,
$\Omega_{8}=     0.942914$,
$\Omega_{10}=    -0.860525$,
$\Omega_{12}=     0.770493$,
$\Omega_{14}=    -0.675846$,
$\Omega_{16}=     0.580283$,
$\Omega_{18}=    -0.486451$,
$\Omega_{20}=     0.397320$,
$\Omega_{22}=    -0.314853$,
$\Omega_{24}=     0.241036$,
$\Omega_{26}=    -0.176940$,
$\Omega_{28}=     0.123418$,
$\Omega_{30}=    -0.080579$,
$\Omega_{32}=     0.048126$,
$\Omega_{34}=    -0.025290$,
$\Omega_{36}=     0.010730$,
$\Omega_{38}=    -0.003034$

$N_\Omega=21$:
$\alpha_{1} = 0.039729$,
$\alpha_{2} = 0.107746$,
$\alpha_{3} = 0.203686$,
$\alpha_{4} = 0.320731$,
$\alpha_{5} = 0.450563$,
$\alpha_{6} = 0.583957$,
$\alpha_{7} = 0.711432$,
$\alpha_{8} = 0.823930$,
$\alpha_{9} = 0.913455$,
$\alpha_{10} = 0.973637$;
$\Omega_0= 1$,
$\Omega_{2}=    -1.103050$,
$\Omega_{4}=     1.077489$,
$\Omega_{6}=    -1.030870$,
$\Omega_{8}=     0.970394$,
$\Omega_{10}=    -0.898899$,
$\Omega_{12}=     0.819660$,
$\Omega_{14}=    -0.735041$,
$\Omega_{16}=     0.647973$,
$\Omega_{18}=    -0.560615$,
$\Omega_{20}=     0.475461$,
$\Omega_{22}=    -0.394267$,
$\Omega_{24}=     0.318906$,
$\Omega_{26}=    -0.250564$,
$\Omega_{28}=     0.190366$,
$\Omega_{30}=    -0.138832$,
$\Omega_{32}=     0.096280$,
$\Omega_{34}=    -0.062563$,
$\Omega_{36}=     0.037213$,
$\Omega_{38}=    -0.019497$,
$\Omega_{40}=     0.008251$,
$\Omega_{42}=    -0.002333$

$N_\Omega=23$:
$\alpha_{1} = 0.033803$,
$\alpha_{2} = 0.092008$,
$\alpha_{3} = 0.174892$,
$\alpha_{4} = 0.277451$,
$\alpha_{5} = 0.393487$,
$\alpha_{6} = 0.515992$,
$\alpha_{7} = 0.637566$,
$\alpha_{8} = 0.750864$,
$\alpha_{9} = 0.849043$,
$\alpha_{10} = 0.926171$,
$\alpha_{11} = 0.977582$;
$\Omega_0= 1$,
$\Omega_{2}=    -1.105308$,
$\Omega_{4}=     1.084451$,
$\Omega_{6}=    -1.044589$,
$\Omega_{8}=     0.992235$,
$\Omega_{10}=    -0.929737$,
$\Omega_{12}=     0.859695$,
$\Omega_{14}=    -0.783977$,
$\Omega_{16}=     0.704930$,
$\Omega_{18}=    -0.624313$,
$\Omega_{20}=     0.544203$,
$\Omega_{22}=    -0.466119$,
$\Omega_{24}=     0.391741$,
$\Omega_{26}=    -0.322219$,
$\Omega_{28}=     0.258733$,
$\Omega_{30}=    -0.201979$,
$\Omega_{32}=     0.152571$,
$\Omega_{34}=    -0.110714$,
$\Omega_{36}=     0.076443$,
$\Omega_{38}=    -0.049493$,
$\Omega_{40}=     0.029346$,
$\Omega_{42}=    -0.015341$,
$\Omega_{44}=     0.006479$,
$\Omega_{46}=    -0.001832$

$N_\Omega=25$:
$\alpha_{1} = 0.029107$,
$\alpha_{2} = 0.079452$,
$\alpha_{3} = 0.151682$,
$\alpha_{4} = 0.242041$,
$\alpha_{5} = 0.345834$,
$\alpha_{6} = 0.457666$,
$\alpha_{7} = 0.571724$,
$\alpha_{8} = 0.682079$,
$\alpha_{9} = 0.782997$,
$\alpha_{10} = 0.869230$,
$\alpha_{11} = 0.936297$,
$\alpha_{12} = 0.980705$;
$\Omega_0= 1$,
$\Omega_{2}=    -1.107092$,
$\Omega_{4}=     1.089990$,
$\Omega_{6}=    -1.055575$,
$\Omega_{8}=     1.009862$,
$\Omega_{10}=    -0.954847$,
$\Omega_{12}=     0.892630$,
$\Omega_{14}=    -0.824713$,
$\Omega_{16}=     0.752995$,
$\Omega_{18}=    -0.678916$,
$\Omega_{20}=     0.604201$,
$\Omega_{22}=    -0.530149$,
$\Omega_{24}=     0.458227$,
$\Omega_{26}=    -0.389494$,
$\Omega_{28}=     0.325078$,
$\Omega_{30}=    -0.265727$,
$\Omega_{32}=     0.212172$,
$\Omega_{34}=    -0.164805$,
$\Omega_{36}=     0.123933$,
$\Omega_{38}=    -0.089584$,
$\Omega_{40}=     0.061641$,
$\Omega_{42}=    -0.039796$,
$\Omega_{44}=     0.023538$,
$\Omega_{46}=    -0.012284$,
$\Omega_{48}=     0.005180$,
$\Omega_{50}=    -0.001465$


\medskip

Various perturbative results for instabilities can be calculated
directly from the $\Omega_l$'s using the following formula for the
transverse gluon self-energy in the special case
that the gluon momentum $\k$ points along the beam axis \cite{linear1}:
\begin {subequations}
\label{eq:Piperp}
\begin {equation}
  \Pi_\perp(\omega, k {\bm e}_z)
  =
  \half m_\infty^2 \sum_\ell \sqrt{2\ell{+}1}\;
  \kappa_\ell\!\left(\frac{\omega}{k}\right) \, {\Omega_\ell}
\end {equation}
with
\begin {equation}
  \kappa_\ell(\eta) \equiv
  (1+\eta^2) \delta_{\ell 0}
  + (1-\eta^2) [(\ell+1) Q_{\ell+1}(\eta) - (\ell-1) \eta Q_\ell(\eta)] .
\end {equation}
\end {subequations}
Here, $Q_l(\eta)$ is the Legendre function of the second kind defined
so that it is regular at $\eta=\infty$ and the cut is chosen to
run from $-1$ to +1.  For example,
$Q_0(z) = \frac12 \ln[(z+1)/(z-1)]$.
The corresponding dispersion relation is
\begin {equation}
   -\omega^2 + k^2 + \Pi_\perp(\omega,k {\bm e}_z) = 0 .
\end {equation}
For a given distribution $\Omega(\theta)$, one can solve this equation
numerically for each $k$, which is how Fig.\ \ref{fig:gamma_vs_k} was
produced.  By scanning over $k$, the largest growth rate
$\gamma = \Im\omega$ and corresponding wavenumber $k_*$ can be
found.
The remaining parameters in Table \ref{tab:fguy} are
given in terms of the $\Omega_l$ as
\begin {equation}
  (v_z)_{\rm rms} = \sqrt{\frac13 
      \left(1 + \frac{2 \Omega_2}{\sqrt{5}}\right)} 
\end {equation}
and \cite{ALM}
\begin {equation}
  \kmax = \left[ - \lim_{k\to 0} \Pi_\perp(0,k {\bm e}_z) \right]^{1/2}
  = \left\{ -
      \half m_\infty^2 \sum_\ell \sqrt{2l{+}1}\;
       \left[
         \delta_{\ell 0} - \frac{(-)^{\ell/2} \, \ell!!}{(\ell-1)!!}
       \right]
        {\Omega_\ell}
    \right\}^{1/2} .
\end {equation}


\section{Damping large \boldmath$\ell$ and \boldmath$m$ modes}
\label {app:damp}

In previous work for moderate anisotropy \cite{linear2}, we found
we could reduce errors from a finite $\lmax$ cut-off by damping
$\ell$ modes near the cut-off.
There, we modified the equations of motion for
the $W_{\ell m}$'s to
\begin{equation}
\frac{dW_{\ell m}}{dt} = (\mbox{original}) - \gamma_{\rm damp} W_{\ell m}
	\, \Theta(\half + \ell - \ldamp) \, ,
\label {eq:old_damp}
\end{equation}
where $\Theta(z)$ is the step function.
This introduced damping for all modes with $\ell$ between $\ldamp$ and
$\lmax$.  We chose
\begin {equation}
  \ldamp = \left\lfloor \tfrac23 \lmax \right\rfloor,
  \qquad
  \gamma_{\rm damp} = \frac{m_\infty}{\sqrt{\lmax}} .
\label {eq:damp_params}
\end {equation}
The rationale behind the choice of $\gamma_{\rm damp}$ is discussed
in Ref.\ \cite{linear2}.  In brief, energy in $W_{\ell m}$'s of large
$\ell,m$ tends to cascade to still higher $\ell,m$.  The presence of a
cutoff can ``reflect'' the energy back to low $\ell,m$, which is
unphysical.  Damping avoids this by absorbing this energy, which
reproduces the physics of its cascading to arbitrarily high $\ell,m$.

In the current work, we introduce similar damping near
the cut-off $\mmax$ on $m$.  Also, we have changed the procedure to
turn on the amount of damping more gradually as $\ell$ or $m$ increase.%
\footnote{
  We thank Michael Strickland for suggesting this improvement.
}
This prevents reflection at the boundary between modes which are and are
  not damped.  In this paper, we replace (\ref{eq:old_damp})
by
\begin{equation}
\frac{dW_{\ell m}}{dt} = (\mbox{original})
  - \gamma_{\ell m} W_{\ell m}
	\, \Theta(\half + \ell - \ldamp) \, \Theta(\half + m - \mdamp) ,
\end{equation}
\begin {equation}
  \gamma_{\ell m}
  = \frac{m_\infty}{\sqrt{\lmax}}
        \frac{(\half+\ell-\ldamp)}{(\lmax-\ldamp)}
  + \frac{m_\infty}{\sqrt{\mmax}}
        \frac{(\half+m-\mdamp)}{(\mmax-\mdamp)}
  \,,
\end {equation}
with
\begin {equation}
  \ldamp = \left\lfloor \tfrac23 \lmax \right\rfloor
  \quad \mbox{and} \quad
  \mdamp = \left\lfloor \tfrac23 \mmax \right\rfloor .
\end {equation}



\end {document}